\pdfoutput=1
\documentclass{article}

\usepackage{arxiv}

\usepackage[utf8]{inputenc} 
\usepackage[T1]{fontenc}    
\usepackage{hyperref}       
\usepackage{url}            
\usepackage{booktabs}       
\usepackage{amsfonts}       
\usepackage{nicefrac}       
\usepackage{microtype}      
\usepackage{amsmath}        
\usepackage{lipsum}		
\usepackage{subcaption} 
\usepackage{graphicx}
\usepackage{natbib}
\usepackage{doi}

\title{Public Goods Game on Complex Networks: the interplay between conformity and topology}

\author{{Ren ~Manfredi}\\
	IMT School for Advanced Studies\\
	Lucca, Italy\\
	\texttt{ren.manfredi@imtlucca.it} \\
	\And
	{Eugenio ~Vicario} \\
	IMT School for Advanced Studies\\
	Lucca, Italy\\
	\texttt{eugenio.vicario@imtlucca.it } \\
    \And
	{Ennio ~Bilancini} \\
	IMT School for Advanced Studies\\
	Lucca, Italy\\
	\texttt{ennio.bilancini@imtlucca.it } \\
    \And
	{Rossana ~Mastrandrea} \\
	University of Turin\\
	Turin, Italy \\
	\texttt{rossana.mastrandrea@unito.it} \\
}

\hypersetup{
pdftitle={A template for the arxiv style},
pdfsubject={q-bio.NC, q-bio.QM},
pdfauthor={Ren Manfredi, Eugenio Vicario, Ennio Bilancini, Rossana Mastrandrea},
pdfkeywords={Cooperation, Evolutionary game-theory, Public Goods Game, Conformism, Network, Agent-based modeling.},
}

\begin{document}
\maketitle
\begin{abstract}
	Human cooperation is a phenomenon that has been extensively studied, and to date several explanations have been proposed, from network reciprocity to behavioral mechanisms that incorporate social and cognitive aspects. In this work, we studied the combined effect of conformity and network structure on the evolution of cooperation in the spatial Public Goods Game. By assigning agents different individual sensitivities to payoffs and neighborhood behavior, we explored the cooperative dynamics of this heterogeneous population on both regular and complex topologies. Our results show how the interaction between conformity and the distinctive features of each network can lead to very different outcomes, from the promotion of cooperation in regular topologies to null or negative effects in heterogeneous networks.
\end{abstract}


\section{Introduction}\label{introduction_ch1}

In his influential article published in 1968, Garrett Hardin metaphorically illustrated the “tragedy of the commons” with the depiction of an open pasture overgrazed by (maximizing) shepherds \citep{hardin1968tragedy}. Since each shepherd is incentivized to increase his flock, the inevitable outcome is the depletion of the common good, a harmful ending for all herdsmen. By seeking to obtain the maximum individual benefit, the shepherds ended up achieving the opposite goal, harming themselves and, ultimately, the entire community.\\
\noindent Despite its simplicity, the social dilemma depicted by Hardin’s metaphor serves as a good illustration of some of the contemporary social challenges, such as climate change or antimicrobial resistance \citep{van2022human, harring2021social}. The dilemma arises from a fundamental conflict between individual benefit and collective interest. The individual payoff from non-cooperative behavior, such as the overuse of antibiotics or the exploitation of shared environmental resources, exceeds that from cooperative actions, regardless of others' behavior. However, the social optimum, in which everyone would be better off, is achieved only if everyone cooperates.\\
\noindent In game theory, this situation is formally described by any game in which at least one equilibrium is Pareto inefficient \citep{archetti2012game}: there is, in fact, an alternative outcome in which an agent could achieve a better result without reducing that of others, but at the same time, no one is incentivized to modify their behavior. Since its first formal appearance in 1944 by \citet{von1944theory}, game theory has proposed numerous mechanisms to explain how cooperation emerges and is sustained over time \citep{traulsen2023future}. Among its most important developments is Evolutionary Game Theory (EGT), whose main contribution has been to adopt a dynamic approach to the study of cooperation, relaxing some of the classical assumptions such as perfect rationality and well-mixed populations. This has made it possible to transition to a more complex and rich representation of human behavior, in which agents - now constrained by finite cognitive resources and incomplete information - do not maximize their utility but instead adopt adaptive (heuristic) decision-making procedures to achieve satisfactory and often suboptimal results \citep{simon1955behavioral, selten1990bounded}. Furthermore, the use of structured populations has made it possible to introduce and represent the structure of interactions among agents, thereby capturing another important feature of the social dimension of human behavior \citep{nowak2010evolutionary, perc2013evolutionary}.\\
\noindent These efforts to outline a more realistic representation of human behavior are also reflected in a recent line of research focused on the update rule, i.e. how agents adapt their strategies and adjust their behavior. The purpose of these studies has been to investigate the effect on cooperation of an update rule that differs from or serves as an alternative to the traditional payoff-based imitation mechanism. The classical approach assumes, in fact, that the main behavioral driver for agents is the maximization of their payoff or, in evolutionary terms, their fitness. One way to implement this dynamic is to assume an imitation process of those strategies that perform better than one’s own. Strategies are therefore subject to a selection process, which may eventually lead to their disappearance or spread within the population.\\
\noindent Although imitation may be regarded as a (cognitive) heuristic that allows us to move beyond the assumption of perfect rationality \citep{sunstein2019conformity}, on the other hand, it does not adequately capture the social dimension of human behavior. Significant inter-individual differences in behavior can, in fact, be attributed not only to mere opportunistic calculations or simple decision-making shortcuts, but also to social factors capable of influencing our actions both implicitly and explicitly \citep{bicchieri2005grammar}. Observable behavior is therefore the synthesis of multiple determinants – social, cognitive, and cultural.\\
\noindent In light of these limitations, a recent line of research has sought to investigate alternative update rules, taking into account one of the major social determinants of behavior, namely conformity. Conformity can be defined as the pervasive behavioral tendency to align oneself with the behavior perceived as being in the majority. It is therefore an individual bias, influenced by different possible social and psychological drivers \citep{d2020social,bicchieri2005grammar,sunstein2019conformity}. Most of the work has focused on frameworks with pair-wise interactions, typically the Prisoner Dilemma (PD), exploring the impact of conformity on both homogeneous and heterogeneous networks. Given the social nature of this new element, the spatial dimension is in fact an almost natural ingredient.
Different mechanisms of conformity have been proposed, and there is no unified framework in the literature. In some models, the population is divided into payoff-driven and conformity-driven agents \citep{szolnoki2015conformity,szolnoki2016leaders, hu2019effect}, while in others, each agent uses both drivers to update their strategy \citep{pena2009conformity, lin2020evolutionary,cui2013impact}. Specific rules have been developed to investigate particular types of conformity. In \citet{huang2023dual}, conformist agents are distinguished as informational or normative conformists, while in \citet{fang2019effect} they are divided into rational or irrational conformists. Although representations of conformity are very heterogeneous, it is still possible, with due caution, to identify a general agreement on its effects on cooperation.\\
\noindent In the case of frameworks such as the PD, the introduction of conformity on regular networks seems to have a overall positive effect on cooperation, enhancing spatial reciprocity. Clusters of cooperators form more easily and become more resilient to invasions by defectors, who are no longer able to effectively erode their boundaries \citep{pena2009conformity,szolnoki2015conformity}. In heterogeneous networks, however, the effects are more complex and conformity can hinder cooperation under certain conditions. Hubs, which previously played a leading role for other nodes thanks to their greater payoff, are now potentially vulnerable to the behavioral trend of their immediate neighbors \citep{pena2009conformity}. This detrimental effect on cooperation in heterogeneous networks can be counteracted under specific conditions, for example when the payoff is degree-normalized \citep{szolnoki2015conformity} or if the tendency to conform is inversely proportional to the degree \citep{szolnoki2016leaders}.\\
\noindent A minority strand of literature has instead investigated the role of conformity in group-interaction frameworks like the Public Goods Game (PGG), almost exclusively on regular networks. In this case too, the representation of conformism is rather heterogeneous. 
In \citet{fang2019effect}, for example, agents are divided into payoff-driven and conformity-driven, the latter in turn distinguished as irrational and rational. The former, who follow the majority only if they are doing better than their neighbors, provide important support for cooperation, allowing many small clusters of cooperators to survive and preventing the formation of a large cluster of defectors. In \citet{javarone2016conformity}, the population is also divided into payoff- and conformity-driven, but the impact of different densities of the two types on cooperation is studied, showing how the presence of conformist agents can promote the formation of ordered phases in the system. In particular, conformity acts as a sort of catalyst, favoring the convergence to full defection or cooperation, to the point of making the payoff irrelevant when the population is almost totally conformist. Finally, in \citet{szolnoki2012wisdom}, conformity is instead an endogenous factor that modulates the probability of imitation by payoff-biased agents. Here, agents decrease their propensity to imitate others if their strategy is consistent with that of the surrounding group. This process supports cooperation, making clusters more resilient thanks to the stability of the central nodes and slowing down the invasion of defectors.\\

\noindent In this work, we joined this stream of literature by employing the spatial Public Goods Game to study the evolution of cooperation, but with three specific distinctions. Firstly, we introduced conformism as a behavioral mechanism that complements, rather than replaces, payoff-based imitation. This choice is motivated by the fact that in most previous studies (e.g., \cite{javarone2016conformity,szolnoki2015conformity,szolnoki2016leaders, hu2019effect}), the population is usually divided into purely payoff-driven and purely conformist-driven agents. In this model, instead, we broke this dichotomy by assuming that behavior results from the bidirectional interplay between intra-individual characteristics and the local social context. 
Secondly, we considered a heterogeneous population of agents. In the minority of studies in which agents' decisions are determined by a mix of individual and social drivers, agents are often homogeneous, sharing the same global (exogenous) conformism parameter (e.g., \cite{pena2009conformity,lin2020evolutionary, yang2017enhancement}). In this model, by contrast, agents differ from one another in terms of the weight they assign to the payoff and to the behavior of others. Finally, with regard to the topologies employed, we explore the evolution of cooperation also on heterogeneous networks (scale-free and small-world), which are often overlooked in works on conformity in spatial PGG. \\
\noindent Our results show how conformity can interact with network-specific characteristics responsible for promoting cooperation, leading to distinct effects for each topology on strategic dynamics and final levels of cooperation. In particular, for regular networks such as circular lattices, conformity exhibits a homogenizing effect that reinforces existing trends driven by payoff and network reciprocity, with positive effects on cooperation. Conversely, in complex networks such as small-world and scale-free topologies, conformity has negligible effects in the former case and a detrimental impact on cooperation in the latter, as it makes influential hubs vulnerable to the behavior of less-connected neighbors, compromising their ability to amplify cooperation.\\

\noindent The paper is structured as follows: Section \ref{model_ch1} describes the model, the mechanism used to represent conformity, and the simulation routine employed. The results, divided by each topology, are reported and discussed in Section \ref{results_ch1}. Following the conclusions in Section \ref{conclusions_ch1}, an appendix (see \ref{appendix}) is provided for further methodological insights.

\section{Model} \label{model_ch1}

\subsection{General configuration}
\label{general_configuration}

\paragraph{Stage game.}In the standard version of the Public Goods Game (PGG), we consider a population of $N \geq 2$ players, each of whom is endowed with an amount $e_i$. Assuming mandatory participation, each individual has two possible strategies: cooperate (C) by providing a donation $c_i^C = e_i$ to the common pool or defect (D) by not contributing ($c_i^D = 0$). The donations collected in the common pool are multiplied by a factor $r$, called the enhancement factor (often referred to as the productivity factor or synergy factor), and then equally distributed among all players, defectors included. The amount $\pi_i$ that agent $i$ receives is therefore given by the donations collected by the fraction of cooperators, multiplied by the enhancement factor and reduced by the cost $c_i$ only if the agent's is a cooperator. Since cooperators bear a fixed cost $c$,  we can see that the social dilemma arises when $r < N$. For such a condition, the payoff of defectors $\pi_D$ will always be greater than that of cooperators $\pi_C$, making defection the dominant strategy and bringing the frequency of cooperators to zero over time. On the other hand, social optimum is achieved when all agents contribute to the common pool, maximizing total group payoff.\\
\noindent Moving to the spatial PGG, the $N$ agents are now placed in a network structure. Each agent plays simultaneously $k+1$ games, where $k$ is the number of its neighbors: one centered on itself and $k$ on its neighbors. Like the standard PGG, at each round, the agent can decide whether to contribute to the common pool or not, keeping the same strategy in all $k+1$ PGGs. All contributions of a single game are then multiplied by the enhancement factor $r$ and equally redistributed among all participants. The payoff $\pi$ for any agent $i$, with a fixed cost per game $c = 1$, is given by:

\begin{equation} \label{payoff1} 
\pi_i = \frac{r (s_i + \sum_{j=1}^{k_i} s_j)}{k_i+1} + \sum_{j=1}^{k_i}\frac{r(s_j +\sum_{z=1}^{k_j} s_z)}{k_j + 1} - (k_i + 1)s_i
\end{equation}

\noindent In Equation \ref{payoff1}, we denote by $k$ the number of neighbors of an agent and by $r$ the enhancement factor. The term $s$ is a dummy variable, which indicates the agent's strategy: if cooperator $s = 1$, otherwise $s = 0$. The terms $j$ and $z$ indicate the neighbors of agent $i$ and the neighbors of each agent $j$, respectively. Consequently, agent $i$ is included among agents $z$, since it is also a neighbor of $j$. The first term of Equation \ref{payoff1} therefore represents the portion of agent $i$'s payoff obtained from the game centered on itself, while the second term represents the payoff obtained by participating in the games of its neighbors. Finally, the last term is the total cost incurred by the agent, only if the latter is a cooperator.

\begin{figure}[hbt]
    \centering
    \includegraphics[width=0.5\linewidth]{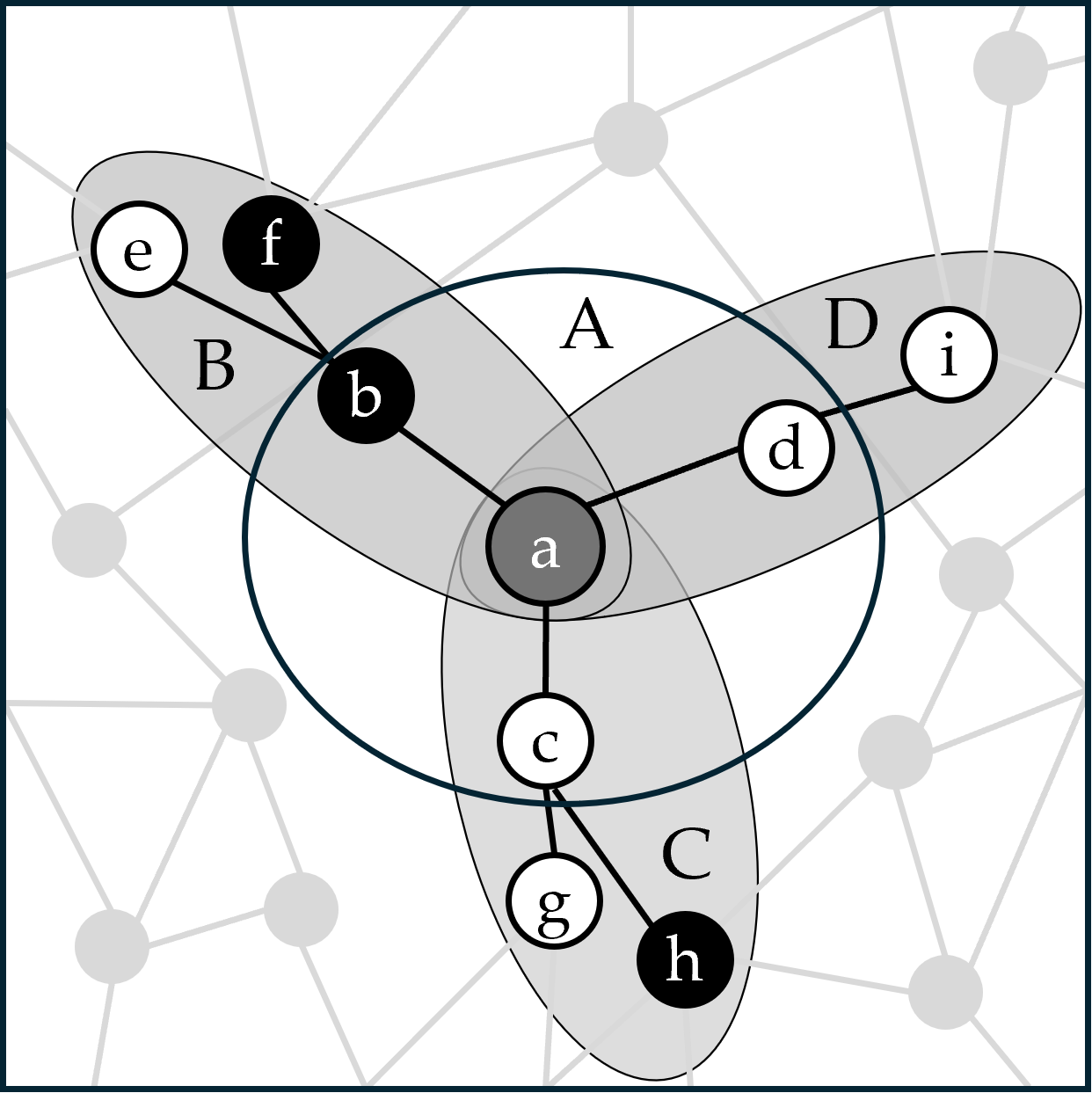}
    \caption{A visual representation of the spatial PGG. The central node $a$ participates in a total of four PGG. The first ($A$) is centered on itself; the other three ($B$, $C$, $D$) are centered on its immediate neighbors $b$, $c$, and $d$, respectively. Adapted from \citet{santos2008social}.}
    \label{fig:SPGG_Example}
\end{figure}

\begin{figure}[hbt]
    \centering
    \includegraphics[width=0.9\linewidth]{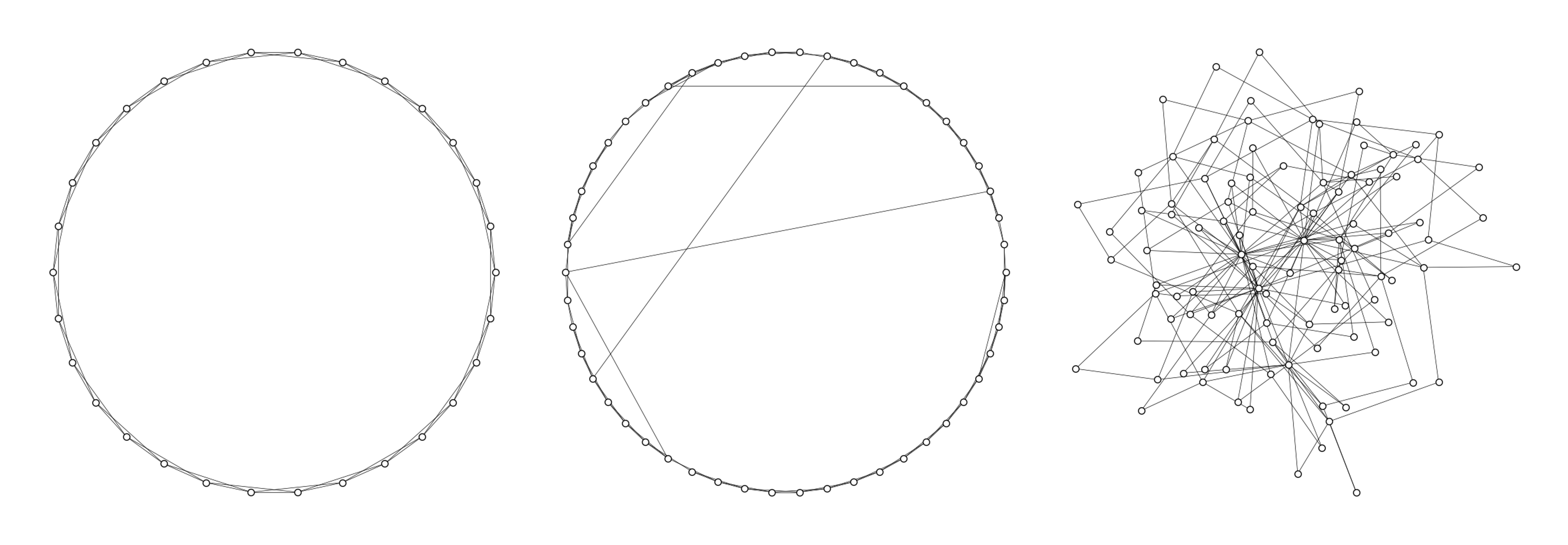}
    \caption{Graphical representation of the three types of networks employed. Starting from the left, circular lattice, small-world, and scale-free.}
    \label{fig:Networks_plot}
\end{figure}

\paragraph{Interaction network.}We consider a population of $N$ agents placed in a network structure, ensuring the absence of isolated nodes. The number of links for each agent that occupies a \textit{node} -- named node \textit{degree} and indicated with $k$ -- represents the number of local PGGs in which the agent participates and plays a key role in shaping their perception of the local social norm. We introduce three different topologies, represented in Figure \ref{fig:Networks_plot}, with average degree $\overline{k} = 4$: (i) regular circular lattice; (ii) small-world network with rewiring probability $p = 0.05$; (iii) scale-free network with attachment parameter $m = 2$ (to ensure an average degree $\overline{k} = 4$). The last two topologies were generated using the Watts–Strogatz model \citep{watts1998collective} and the Barabási and Albert model \citep{barabasi1999emergence, barabasi1999mean}, respectively.\\
\noindent The three population structures are employed to explore how network properties can influence cooperation and modulate the effect of conformism. The circular lattice is a homogeneous network where all nodes have the same degree $k$. It is a very sparse, highly clustered, one-dimensional regular structure with a large average path length, which increases as $N$ increases \citep{wang2003complex}. The small-world is also considered a homogeneous network, where all nodes have approximately the same degree $k$. Since it can be considered as an interpolation between regular and random networks, this topology exhibits a Poisson-like degree distribution and shares the high clustering property of lattices, while manifesting shorter average path lengths. These last two properties are present when the rewiring probability $p$ is within the range $0.001‹p‹0.1$ \citep{watts1998collective}. For this reason, in this work we employ $p = 0.05$. Finally, scale-free networks are characterized by a heterogeneous degree distribution: a few hyperconnected nodes (hubs) dominate the network, while the majority of nodes have a low degree, usually below the average $\overline{k}$. Compared to the random network, scale-free exhibit a smaller average path length and a higher clustering coefficient \citep{wang2003complex}.

\subsection{Dynamics}
At the end of each round, all agents have the opportunity to update their strategy through a process of imitation. Each agent $i$ randomly selects a neighbor $j$ from their closest neighborhood, observing its payoff $\pi_j$ and strategy $s_j$. The imitation mechanism consists of two combined rules: payoff-based imitation and conformity. Regarding the former, following the work of \cite{santos2008social}, the probability that agent $i$ will adopt the strategy of neighbor $j$ is given by:

\begin{equation}\label{updatev0}
P_\pi(i \rightarrow j)=  \Big(\frac{\pi_j-\pi_i}{M} \Big)\phi_{\Delta\pi}
\end{equation}

\noindent where $\pi_j ,\pi_i$ represent, respectively, the payoffs of agents $j$ and $i$; $M$ stands for the maximum possible payoff difference between $i$ and all its neighbors (ensuring proper normalization) and $\phi_{\Delta\pi}$ represents a function depending on the payoff difference $\Delta \pi=\pi_{j}-\pi_{i}$:
 \begin{equation}
 \phi_{\Delta\pi}=
 \begin{cases}
0 & \text{if} \quad \Delta_{\pi} \leq 0\\
1 & \text{if} \quad \text{otherwise}
\end{cases}
\end{equation}

\noindent ensuring that the update takes place only if $\pi_i < \pi_j$. If this condition is not met, the update process does not take place and the probability of imitation based on payoff is zero. \\
\noindent The second part that completes the update process is the conformity mechanism. In addition to the probability of imitation driven by the payoff, agent $i$ also considers $j$'s strategy $s_j$ and compares it with that of its neighborhood (agent $i$ included), extrapolating the perceived local social norm $SN_i$, represented by the fraction of defectors, as shown in Equation \ref{SN}.

\begin{equation}\label{SN}
    SN_i= 1 - \frac{s_i + \sum_{j=1} ^{k_i} s_j} {k_i + 1}
\end{equation}

\noindent In previous works (e.g., \cite{javarone2016conformity, szolnoki2015conformity}), these two update mechanisms - payoff and conformism-based imitation - are kept separate, generating a population composed of two types of agents. Conversely, in this model, agents employ both rules, as we assume that both payoff and others' behavior influence the behavior of the individual agent. An important implication of this approach is that agents have the opportunity to imitate their neighbors even when $\Delta \pi \leq 0$. To represent the combined impact of the two mechanisms, each agent $i$ is assigned a parameter $\alpha_i \in (0,1)$, which represents the decision weight of the payoff comparison for agent $i$. Consequently, the value $1 -\alpha_i$ indicates the agent's sensitivity to the behavioral norm perceived from their neighborhood. The resulting final imitation probability is outlined in Equation \ref{updatev1}.\\

\begin{equation}\label{updatev1}
P(i \rightarrow j)= \alpha_i \Big(\frac{\pi_j-\pi_i}{M} \Big)\phi_{\Delta\pi} + (1-\alpha_i)(|SN_i-s_j|)
\end{equation}

\noindent The parameter $\alpha$ is randomly extracted from a Beta distribution for each individual agent during initialization, and the value remains constant and unchanged over time. The result is therefore a real heterogeneous population of agents exhibiting different biases towards payoff and conformity, unlike the majority of the literature, where, regardless of the mechanism employed to represent conformity, agents are always identical to each other.\\

\begin{figure}[hbt]
    \centering
    \includegraphics[width=0.7\linewidth]{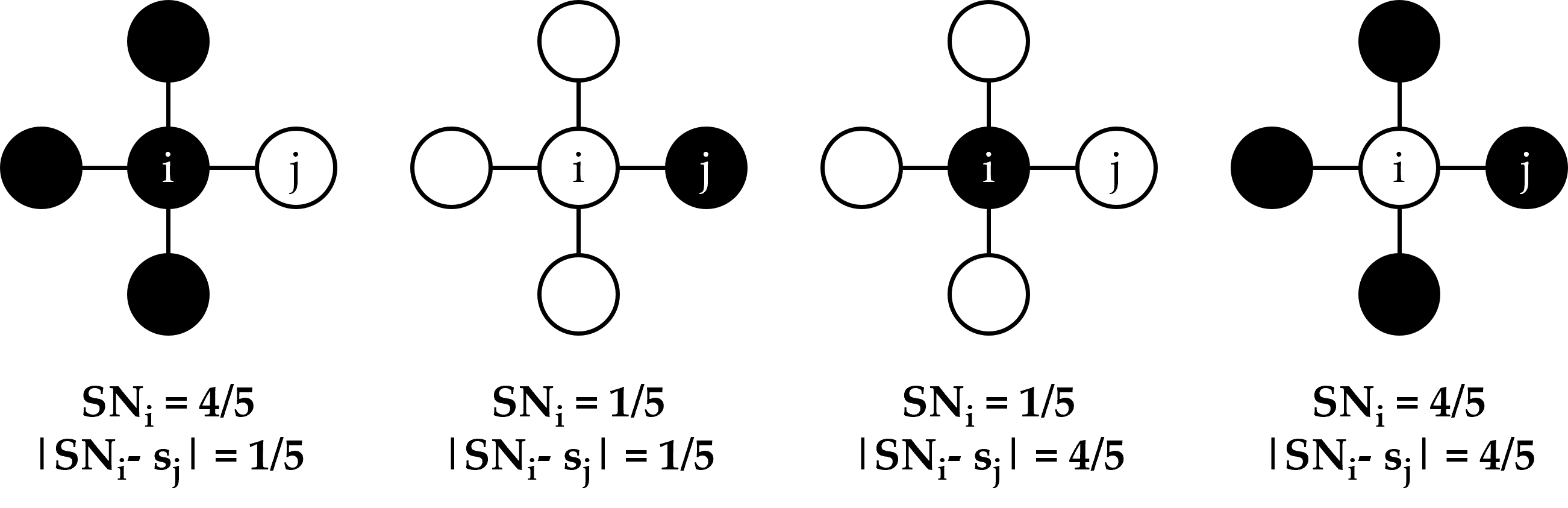}
    \caption{The four different scenarios illustrate the impact of the social norm on agent $i$, who is considering whether to imitate its neighbor $j$. The black circles represent defectors ($s = 0$), whilst the white circles indicate cooperative agents ($s = 1$). The value of $SN_i$ represents the social norm as perceived by agent $i$, whilst the term $|SN_i - s_j|$ represents the influence that neighbor $j$ exerts on $i$. For example, in the first case on the left, $j$ is the only agent to adopt a strategy different from the rest of the neighborhood, and therefore exerts minimal influence on $i$ ($|SN - s_j| = 1/5$).}
    \label{fig:SN_EqExample}
\end{figure}

\noindent It is worth noting three aspects of the conformism mechanism. First, agent $i$ evaluates the social conformity of the agent $j$ strategy to decide whether to imitate or not. During this evaluation, the influence exerted by $j$ on $i$ is modulated by the level of consistency with the local norm, as shown in Figure \ref{fig:SN_EqExample}. This implies that even a minority strategy is capable of influencing the decisions of agents, albeit to a lesser extent than a majority strategy. Second, the social norm is neither positively nor negatively characterized, but it drives agents to conform to their neighborhood, regardless of the type of strategy (C or D). Third, network structure plays an important role in the conformity-based imitation mechanism, as it affects agent's social awareness and the extent of social information.

\subsection{Simulations}\label{Simulations} Simulations were run for a population of $N = 1000$ agents with average connectivity $\overline{k} = 4$. The choice of these parameters is intended to follow the same setting as adopted by \cite{santos2008social}, providing an initial element of comparison. Cooperators initially represent the $50\%$ of the whole population and are randomly distributed on the network structure. The network structure remains unchanged throughout the simulations and all agents update their strategy synchronously at each time step.
For each value of the enhancement factor $r \in \{1,2,3,4,5\}$ we run a total of $10^7$ generations and averaged the results over $300$ initializations ($30$ runs for $10$ different realizations of the same class of graph). Each simulation may end before reaching the maximum number of iterations if, after $10^5$ time steps, the average value of cooperation remains unchanged for 30 consecutive iterations, or if the system is completely taken over by one of the two types of agents, cooperators (C) or defectors (D).\\
\noindent The parameter $\alpha_i$ was drawn from a Beta distribution independently for each agent and kept fixed through all simulations. The Beta distribution is a continuous probability distribution bounded in the interval $(0,1)$ and defined by the following probability density function:

\begin{equation} \label{beta_dist}
    B(x; a,b) = \frac{x^{a-1}(1-x)^{b-1}}{\beta(a,b)}, \quad 0 < x < 1
\end{equation}

\noindent By varying the shape parameters $a$ and $b$ ($a,b \geq 1$), we were able to generate heterogeneous populations qualitatively different in terms of susceptibility to payoff and conformity. In particular, we tested the following cases:
\begin{itemize}
    \item $a = b = 1$: the parameter $\alpha$ follows a uniform distribution where each value is equally probable, generating a balanced population with high diversity. We will refer to this case as the Beta(1,1) distribution.
    \item $a < b$: the parameter $\alpha$ follows a skewed right distribution, generating a population with a high propensity for conformity and a minority of agents who are more attentive to payoff. In this work, we explored this scenario through a Beta(1,10) distribution.
    \item $a > b$: the parameter $\alpha$ follows a left-skewed distribution, outlining a population composed of a majority of agents who are more sensitive to payoff than to the behavior of their neighbors. We explored this case through a Beta(10,1) distribution.
\end{itemize}

\noindent In addition to the above distributions, we studied the standard spatial PGG without conformity ($\alpha_i = 1, \forall i$), which we will refer to in the following sections as “NoSN.”

\section{Results}
\label{results_ch1}

\subsection{Lattice}

\begin{figure}[hbt]
    \centering
    \includegraphics[width=0.5\linewidth]{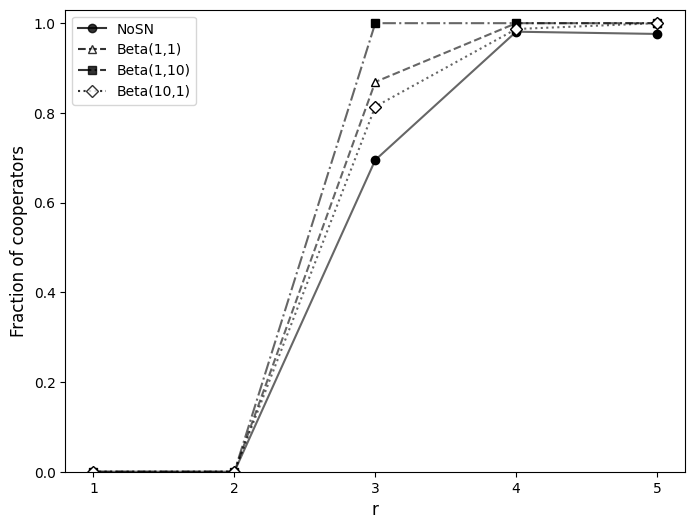}
    \caption{Evolution of cooperation in the circular lattice across the four scenarios considered for different values of the enhancement factor $r$.}
    \label{fig:CL_Overview}
\end{figure}

\begin{figure}[hbt]
    \centering
    \includegraphics[width=0.5
    \linewidth]{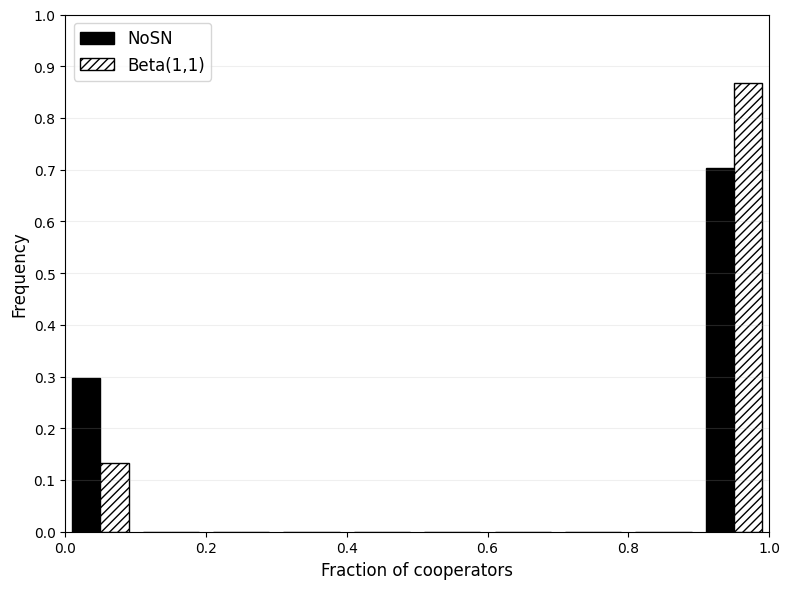}
    \caption{Convergence patterns for $r = 3$ for the NoSN and the Beta(1,1) scenarios, in the circular lattice.}
    \label{fig:CL_Convergence_NoSNvsBeta11}
\end{figure}

\noindent We begin by considering the results of the PGG in the circular lattice.
Figure \ref{fig:CL_Overview} shows the evolution of cooperation as the enhancement factor $r$ and the parameters $(a,b)$ of the Beta distribution vary. In the absence of conformity, cooperators die out for $r \leq 2$, and the system converges to a state of total defection. For higher values of $r$, the payoff gap between defectors and cooperators narrows. Under these more favorable conditions, clusters of cooperators have the opportunity to form. Here, cooperators can survive exploitation by defectors, particularly those agents far from the edges of the cluster. By interacting only with other cooperators, the resulting payoff is sufficient to enable them to survive and, eventually, even invade the system. It is worth noting that, for the values of $k$ and $r$ considered here, the cooperative strategy is never the best reply. However, for $r = 3$, the system alternates between states of total defection and states of high values of cooperation, never reaching full convergence in the latter case. These dynamic disappear for $r \geq 4$, where the cooperative component is strongly predominant but never total. It is easy to show how a hard core of defectors is always able to persist (see Appendix \ref{Appendix_A}); in particular, for the values of $r$ tested, a single defector or a pair of adjacent D agents surrounded by cooperators will always have a zero probability of imitating a C agent, since their payoff will always be higher. Isolated defectors therefore thrive and a total invasion by cooperators is impossible. However, as defectors grow around this hard core, their fitness decreases, making cooperation more advantageous and pushing defection back into a minority position.\\
\noindent The presence of an immovable core of defectors, even for high values of $r$, disappears when the mechanism of conformity is added to the agents' decision-making process. Even a weak presence (e.g., Beta(10,1) distribution) is sufficient to cause the system to eventually converge to total cooperation for values of $r \geq 4$. In fact, while isolation provides defectors with a significant advantage in terms of payoff, it also entails a strong disadvantage in social terms, compromising the defectors' immunity to the cooperative strategy. When the influence of conformity is more balanced, i.e., Beta(1,1), the system converges to total defection for $r \leq 2$ and total cooperation for $r \geq 4$. For the intermediate case $r=3$, convergence can be observed in both directions, but with a more marked tendency toward full cooperation, as shown in Figure \ref{fig:CL_Convergence_NoSNvsBeta11}. If we increase the weight of conformity on the agents' decision-making process, this bistability is reduced until it disappears for the case Beta(1,10). In this scenario, for $r \geq 3$, the cooperators always manage to invade the system sooner or later. \\

\noindent In conclusion, in a regular structure such as the circular lattice, conformity seems to act as a catalyst, reinforcing the convergence of the system towards the most widespread strategy. For $r\leq2$, the environment is hostile to cooperators and the system therefore accelerates towards full defection; the trend is reversed for $r\geq 3$, where conformity pushes more towards full cooperation. This is particularly true as the pervasiveness of conformity in the agents' decision-making process increases: its homogenizing effect reduces the bistable tendency of the system, i.e., the tendency to converge either toward full cooperation or full defection, favoring convergence towards a single direction. The benefit brought to cooperation is mainly represented by the fact that strategically homogeneous clusters exhibit greater resistance to invasions due to conformism and compromise the advantage of isolated defectors given by the payoff.\\

\subsection{Small-world}

\begin{figure}[hbt]
    \centering
    \includegraphics[width=0.5\linewidth]{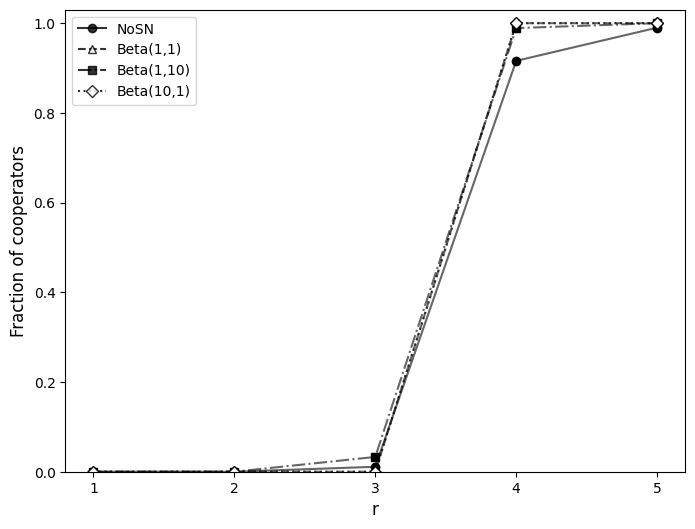}
    \caption{Evolution of cooperation in the small-world across the four scenarios considered for different values of the enhancement factor $r$.}
    \label{fig:SW_overview}
\end{figure}

\noindent As discussed in the previous section, the small-world can be considered a homogeneous network that shares some properties with the regular lattice. In the \cite{watts1998collective} model, this topology is in fact generated starting from a nearest-neighbor coupled network, whose edges are then randomly rewired with a probability $p$. The average number of neighbors $\overline{k}$ is still the same, but the average degree of separation between nodes is now lower \citep{wang2003complex}. Although it may seem like a minor variation, we can observe a significant effect in terms of cooperation.\\
\noindent Figure \ref{fig:SW_overview} shows the final average cooperation as $r$ varies in the small-world network. When the strategy update process is solely payoff-based, the system always converges to total defection for $r \leq 2$, as in the circular lattice. For $r=3$, however, defection remains still the dominant strategy, with rare cases in which a tiny minority of cooperators manages to survive. Only by further increasing the enhancement factor ($r \geq 4$) cooperation reaches a level similar to those observed in the circular lattice. This lag in cooperation is due to the new configuration of each agent's neighborhoods, which no longer correspond, in some cases, to the closest connections as in the regular lattice. This seems sufficient to allow defectors to invade these less compact clusters more easily, requiring a higher enhancement factor $r$ for cooperation to catch up \citep{zhang2017emergence}.\\

\begin{figure}[hbt]
    \centering
    \includegraphics[width=0.5\linewidth]{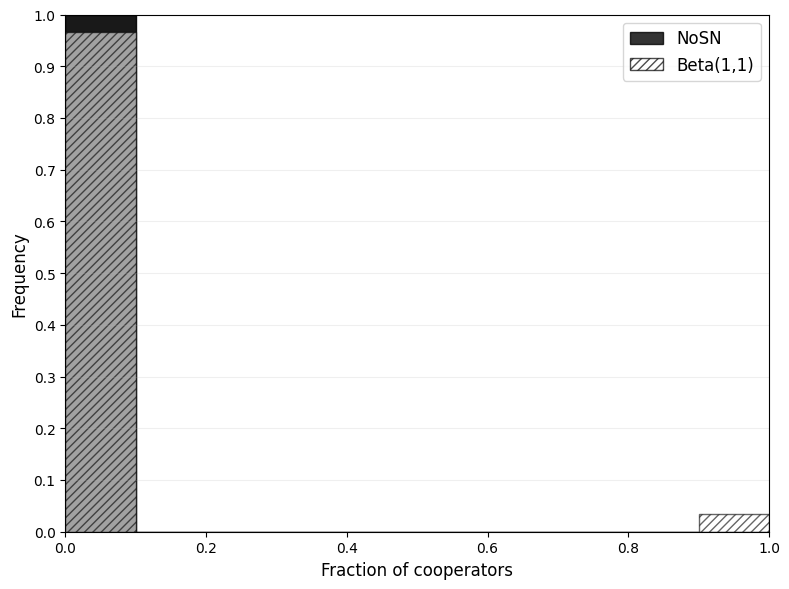}
    \caption{Convergence patterns for $r = 3$ for the NoSN and the Beta(1,1) scenarios, in the small-world.}
    \label{fig:SW_Convergence_NoSNvsBeta110}
\end{figure}

\noindent On a first look, this scenario do not seem to vary significantly with the introduction of conformity. Comparing the different distributions tested, the final average cooperation levels are virtually unchanged (Figures \ref{fig:SW_overview}). Although similar, the results are qualitatively different, as we can see, for example, from the convergence patterns shown in Figure \ref{fig:SW_Convergence_NoSNvsBeta110}. When the role of conformity is weak or moderate, the system evolves in a similar way to when agents are purely payoff-driven. Here, conformity merely homogenizes the agents' strategies. For $r=3$, pockets of resistance among cooperators are definitively invaded by defectors. Similarly, for $r \geq 4$, the system converges to total cooperation.\\
\noindent By further increasing the weight of conformity in the agents' decision-making process, a bistable dynamic emerges. As we have seen, without conformity, cooperators consistently constitute the minority for $r = 3$ and the majority for $r = 4$. In the Beta(1,10) case, however, cooperators manage to completely invade the system for $r = 3$, albeit in rare cases. Similarly, defectors are able to do the same for $r = 4$, again with very limited frequency. \\ 
\noindent In the small-world topology cooperation seems to be disadvantaged by the network structure, compared to the regular lattice. When the decision-making mechanism is purely payoff-based, the dynamics are well defined: for $r \leq 3$, defection is dominant, while for $r\geq 4$, cooperation systematically prevails almost entirely over the system. The moderate presence of conformity merely reinforces this trend. Only when its role becomes more pronounced is it possible to observe a different behavior. For intermediate values of the enhancement factor, this bistability induced by conformity allows minority agents to invade the system, although rarely.

\subsection{Scale-free}
Unlike circular lattices, the different distribution of degrees and the presence of hubs that characterize scale-free networks usually seem to favor cooperation. In the work of \citet{santos2008social}, the authors show how in a heterogeneous topology such as a scale-free network, cooperators are able to spread more quickly than in a regular network, requiring a lower enhancement factor value. The evolutionary advantage conferred by social heterogeneity is mainly represented by hubs, especially in a fixed cost per game regime. Agents occupying hubs tend to play a very high number of games with neighbors that are typically peripheral and with a lower degree, leading to a significant payoff difference in favor of the former. However, the implications and effects on its fitness vary depending on the strategy of its occupant. When defectors occupy the center of the hub, they encourage the spread of defection, which, however, reduces their individual fitness; on the contrary, a cooperating hub increases its fitness as the number of cooperators increases, triggering a virtuous circle. In the latter case, cooperators are able to secure hubs, while a minority of defectors with low degree may still survive.\\

\begin{figure}[hbt]
    \centering
    \includegraphics[width=0.5\linewidth]{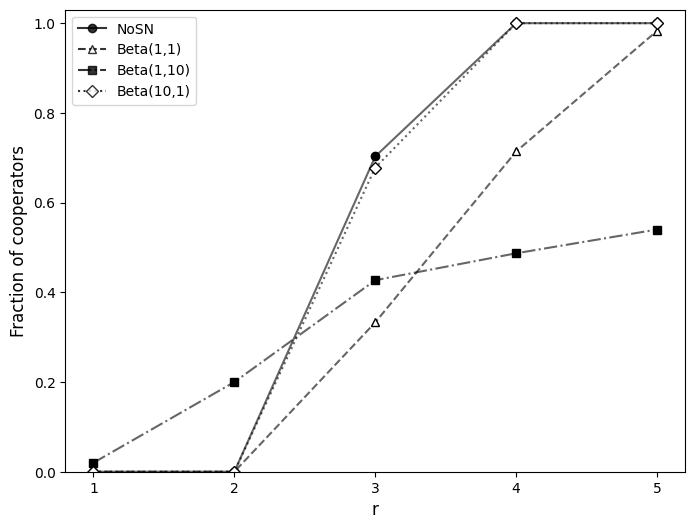}
    \caption{Evolution of cooperation in the scale-free across the four scenarios considered for different values of the enhancement factor $r$.}
    \label{fig:SF_overview}
\end{figure}

\begin{figure}[hbt]
    \centering
    \includegraphics[width=0.5\linewidth]{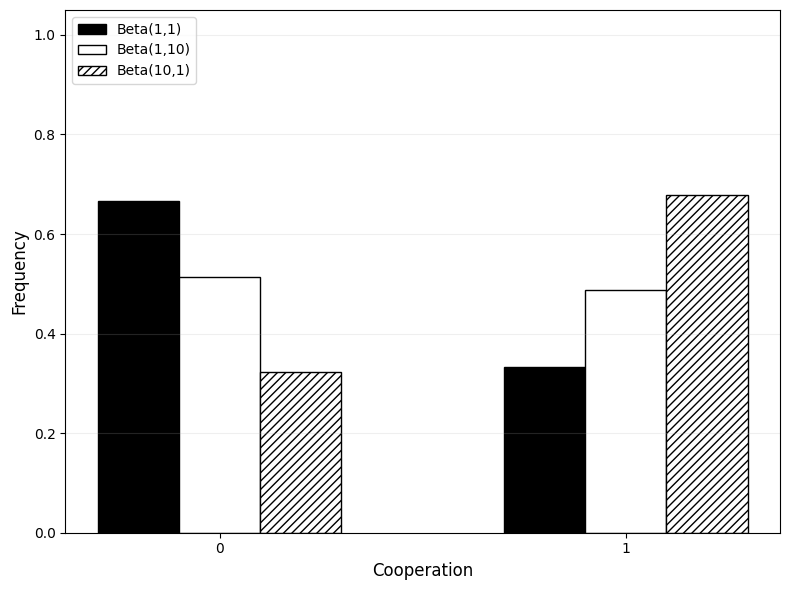}
    \caption{Convergence patterns in the scale-free network for $r = 3$ across the three scenarios with conformism.}
    \label{fig:SF_Convergence}
\end{figure}

\noindent Figure \ref{fig:SF_overview} shows the evolution of cooperation in the scale-free network as the enhancement factor $r$ varies. Consistent with the results of \citet{santos2008social}, cooperation is positive for $r \geq 3$, although the difference observed by the authors in the circular lattice tends to disappear for longer simulation times. In particular, for $r = 3$, the system alternates between states of complete defection and high levels of cooperation, but never total. For $r \geq 4$ cooperators manage to completely invade the system. \\ 
\noindent By introducing the effect of conformity into the population, we can observe the emergence of a bistable dynamic in the system and a general detrimental effect on the evolution of cooperation. Figure \ref{fig:SF_Convergence} shows how the possible final states, for each level of conformity tested, are limited to total defection or total cooperation. The respective average cooperation values shown in Figure \ref{fig:SF_overview}, excluding the case without conformity, can therefore be interpreted as the probability that the system will converge to full cooperation.\\
\noindent As the impact of conformity on the agent' decision-making process increases, cooperation moves away from the results of Santos et al. In the uniform scenario Beta(1,1) cooperation is significantly reduced for $r = 3$ and exhibits values similar to the case without conformity only for $r = 5$. The nature of this dynamic is closely linked to the role played by hubs. In fact, with the introduction of conformity, the evolutionary advantage conferred by the payoff to hubs becomes now fragile. When the imitation process is solely payoff-based, hubs are unlikely to imitate a neighbor with a low degree, given the difference in payoff. However, with the introduction of conformity, hubs now become vulnerable to the majority behavior of their neighbors, significantly increasing the likelihood of changing strategy.\\ 
\noindent By further increasing the weight of conformity on the agents' decision-making process, as in the case Beta(1,10), the role of payoff is even more marginal, leading to a particular effect on the cooperative dynamic. For $r \leq 2$, we now observe positive values of cooperation where they were previously zero, while for $r \geq 3$, cooperation settles at around 50\% on average and appears to be relatively unresponsive to increases in $r$. Low sensitivity to payoff therefore reduces the effect of the evolutionary disadvantage for cooperators usually present for $r \leq2$, but on the other hand, it weakens the tendency to cooperate when materially advantageous ($r \geq 3$).\\ 
\noindent The results therefore reveal a non-trivial effect of conformity on a heterogeneous network such as scale-free networks. Hubs are no longer impregnable fortresses capable of driving their neighborhood, but rather unstable hyperconnected nodes that are vulnerable to the strategic trend of their close connections. Compared to the circular lattice, the homogenizing effect of conformity is stronger, with a greater tendency toward bistability. Finally, high levels of conformity and the resulting lower sensitivity to payoff, promote cooperation in usually unfavorable conditions and vice versa, a phenomenon that we do not observe in regular topologies.

\subsection{Effect of conformity and topology on cooperation}

\begin{figure}[hbt]
    \centering
    \includegraphics[width=0.5\linewidth]{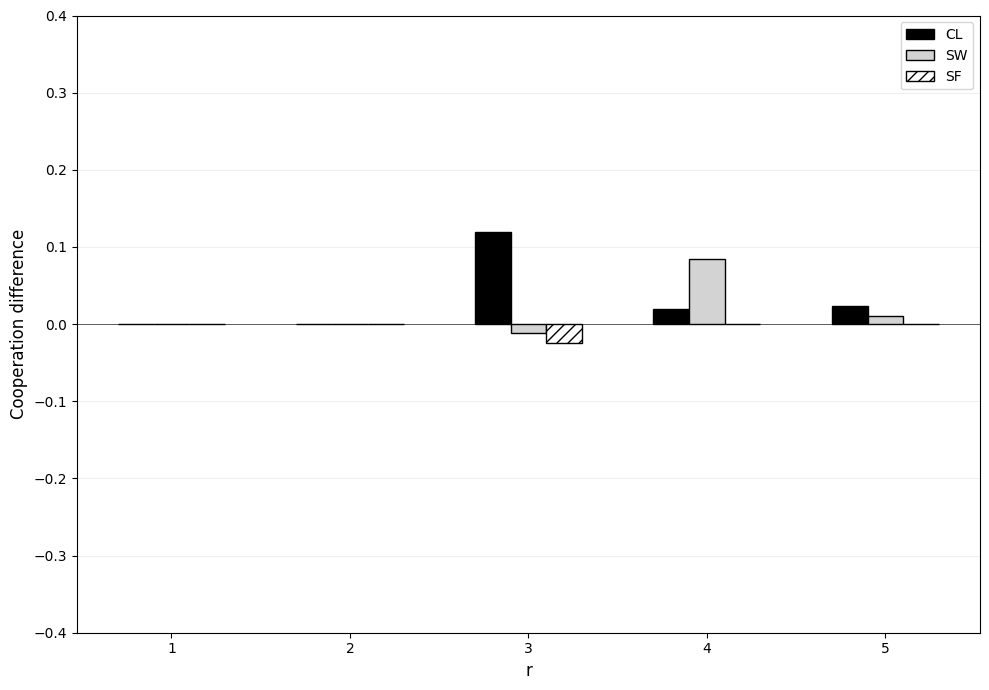}
    \caption{For each of the three topologies employed, the figure shows the cooperation levels for the Beta(10,1) case, expressed as the difference relative to the NoSN scenario.}
    \label{fig:SN_effect_overview_B101}
\end{figure}

\begin{figure}[hbt]
    \centering
    \includegraphics[width=0.5\linewidth]{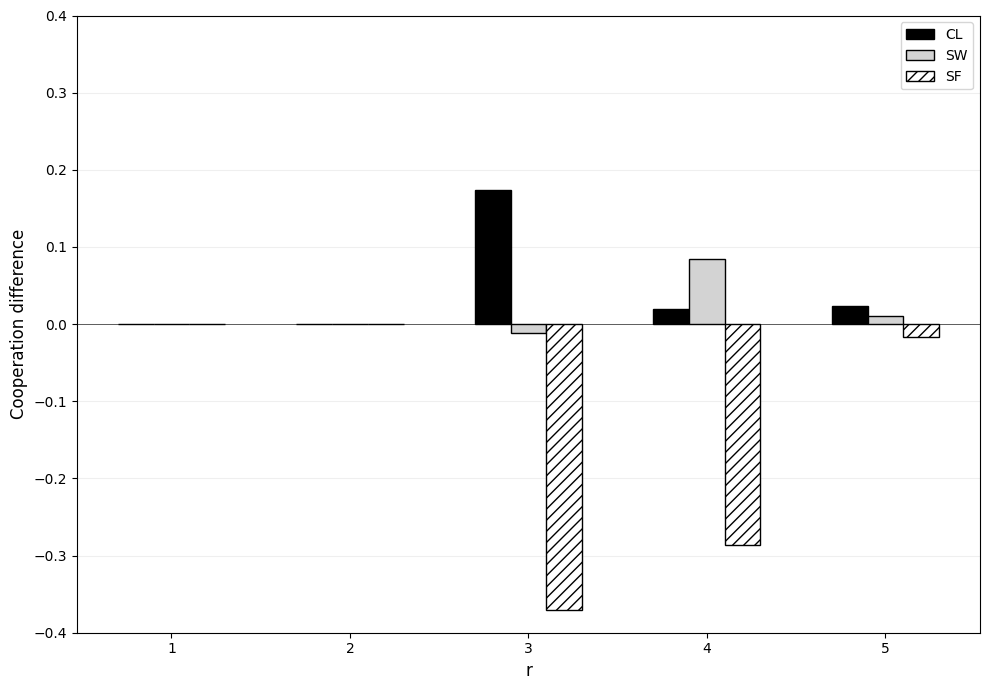}
    \caption{For each of the three topologies employed, the figure shows the cooperation levels for the Beta(1,1) case, expressed as the difference relative to the NoSN scenario.}
    \label{fig:SN_effect_overview_B11}
\end{figure}

In the previous paragraphs, we described the results obtained for the three networks employed and discussed the effect of conformity in each of them. When the behavior of one's neighborhood becomes an important component in each agent's update process, the features of each topology play an important role, influencing the evolutionary dynamics and the final levels of cooperation. 
The homogenizing effect of conformity seems to favor clusters typical of regular networks such as circular lattices. Conversely, in the “weaker” clusters of small-world networks, its effect does not result in significant changes, with average cooperation levels remaining virtually unchanged. In a heterogeneous network such as a scale-free network, conformity triggers strong bistability, and the hubs characteristic of this topology lose their leading role in guiding low-degree agents, with a overall detrimental effect on cooperation.
As we have seen, this interaction between network properties and conformity has distinctive repercussions on the spread of cooperation.\\
\noindent When agents' sensitivity to conformity is weak, as in the case of Beta(10,1), the final levels of cooperation do not differ significantly from the standard case in the absence of conformity (Figure \ref{fig:SN_effect_overview_B101}). Agents are still very sensitive to payoff, and conformity merely reinforces the trend already determined by the latter. However, in the circular lattice, this is sufficient to increase cooperators for $r \geq 3$, ousting the minority of isolated defectors. When conformity assumes greater relevance and is evenly distributed across the population, we observe three different effects on the topologies considered, as shown in Figure \ref{fig:SN_effect_overview_B11}. 
\noindent In the circular lattice, this level of conformity allows the emergence of a bistable dynamic, particularly noticeable when $r = 3$, and the achievement of an ordered phase.\\

\begin{figure}[hbt]
    \centering
    \includegraphics[width=0.5\linewidth]{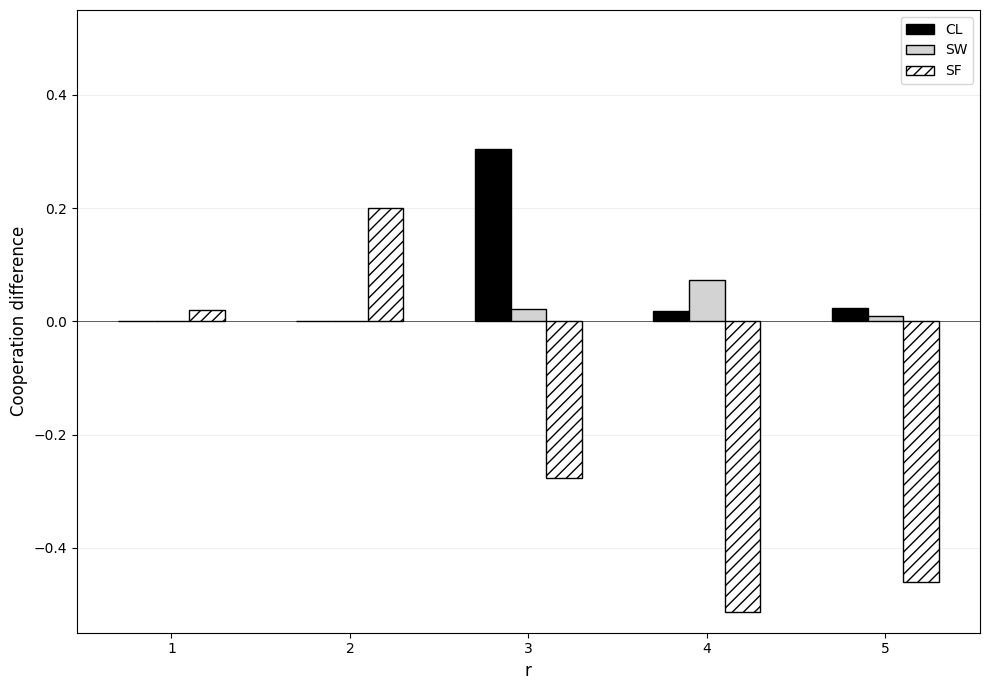}
    \caption{For each of the three topologies employed, the figure shows the cooperation levels for the Beta(1,10) case, expressed as the difference relative to the NoSN scenario.}
    \label{fig:SN_effect_overview_B110}
\end{figure}

\noindent Isolated defectors who survived by exploiting nearby cooperators lose their immunity given by the payoff and become cooperators. On the other hand, the small-world does not exhibit any differences compared to the case Beta(10,1), while in the scale-free network a detrimental effect on cooperation is observed. Sensitivity to the behavior of others seems sufficient to cause hubs to become strategically unstable. 
\noindent A further increase in the role of conformity in updating the strategy confirms the positive effect on cooperation given by homogenization in the case of the circular lattice (Figure \ref{fig:SN_effect_overview_B110}). The cooperative trend present for $r\geq3$ is in fact maximized, as is total defection for $r\leq2$. In contrast, in the small-world, the greater weight of conformity leads to the emergence of bistable behavior, albeit very unbalanced. The payoff here is still decisive in defining the general direction of convergence, but strong conformity can occasionally reverse it. However, it is in the scale-free network that we observe the most peculiar effects of conformity. While in other topologies the payoff and enhancement factor $r$ are generally good predictors of cooperation, the scale-free network proves to be the most sensitive to conformity. As the role of the latter increases, cooperation decreases further and appears relatively unresponsive to increases in the reward $r$. This promotes defection for $r \geq 3$, but at the same time encourages cooperation for $r \leq 2$.

\section{Conclusions}\label{conclusions_ch1}
In this work, we explored the evolution of cooperation through a spatial PGG in which both payoffs and conformity shape the decisions of a heterogeneous population of agents. Using three different topologies, we showed how population structure can modulate the effect of conformity, leading to non-trivial network-specific effects on the final levels of cooperation and on the strategic dynamics of agents. When conformity was included in the agents' decision-making process, we observed a positive impact on cooperation in the circular lattice, an almost neutral effect in the small-world network, and a negative effect in the scale-free network. \\
\noindent In general, in both homogeneous and heterogeneous networks, conformity induces a homogenizing effect in the system, allowing the achievement of a final ordered phase where all agents exhibit the same strategy. For example, in regular networks this occurs because conformity nullifies the immunity guaranteed by the payoff to isolated defectors, who typically thrive by exploiting their cooperative neighbors.\\
\noindent The second effect of conformity was the promotion of a bistable dynamic - often biased - in the three topologies, but with some differences. When the weight of conformity in the strategy update process was moderate, two possible final states could be observed in the circular lattice. This also occurred in the small-world, but only when the role of conformity was strong. In the scale-free, on the other hand, bistability was omnipresent, regardless of the intensity of conformity. Similar populations - in terms of sensitivity to payoff and the behavior of others - therefore exhibited different behaviors based on the topology and structure of interactions in which they operate.\\
\noindent Finally, in regular networks, conformity seemed to reinforce the strategic trends induced by the payoff, strengthening both cooperation and defection based on the enhancement factor value. In heterogeneous networks, on the other hand, cooperation was generally disadvantaged when conformity was moderate or strong. In the latter case, however, it emerged that the heterogeneous topology was more sensitive to conformity than regular networks. Here, payoff seemed to play a particularly marginal role, promoting the emergence of cooperation in usually adverse contexts, but at the same time limiting the impact of favorable conditions on its spread.\\
\noindent In conclusion, we have seen how conformity can influence cooperation, with distinct effects depending on the structure of the population and its features. Our results show that conformity can interact with the structural characteristics responsible for promoting cooperation, from circular lattice clusters to scale-free hubs, with very different effects on cooperation, depending also on the composition of the population.\\
\noindent Finally, in this work, we wanted to apply the spatial PGG framework to complex networks that are usually less explored in the PGG literature, such as small-world and scale-free networks. Furthermore, we employed a truly heterogeneous population of agents, assigning them individual values rather than relying on population-wide parameters or categories. However, we believe that this attempt at greater realism is not yet exhaustive and, like all models, our work is not without limitations. Human beings are part of a constantly evolving social context, where the actions of individuals can alter the latter and vice versa. Opinions and attitudes change over time, through learning mechanisms and interaction with one's environment. The structure of relationships is also dynamic: new links can form, while others can end. Furthermore, each individual can belong to different social groups, in which their role can be central or peripheral. Possible extensions of our model could therefore include the use of dynamic networks or multiplex networks (e.g. \cite{tomassini2020public, wang2023public}), as well as agents capable of updating their characteristics endogenously, in response to the environment or the behavior of others.


\begin{thebibliography}{}

\bibitem[\protect\citeauthoryear{Archetti and Scheuring}{Archetti and Scheuring}{2012}]{archetti2012game}
Archetti, M. and I.~Scheuring (2012).
\newblock Game theory of public goods in one-shot social dilemmas without assortment.
\newblock {\em Journal of theoretical biology\/}~{\em 299}, 9--20.

\bibitem[\protect\citeauthoryear{Barab{\'a}si and Albert}{Barab{\'a}si and Albert}{1999}]{barabasi1999emergence}
Barab{\'a}si, A.-L. and R.~Albert (1999).
\newblock Emergence of scaling in random networks.
\newblock {\em science\/}~{\em 286\/}(5439), 509--512.

\bibitem[\protect\citeauthoryear{Barab{\'a}si, Albert, and Jeong}{Barab{\'a}si et~al.}{1999}]{barabasi1999mean}
Barab{\'a}si, A.-L., R.~Albert, and H.~Jeong (1999).
\newblock Mean-field theory for scale-free random networks.
\newblock {\em Physica A: Statistical Mechanics and its Applications\/}~{\em 272\/}(1-2), 173--187.

\bibitem[\protect\citeauthoryear{Bicchieri}{Bicchieri}{2005}]{bicchieri2005grammar}
Bicchieri, C. (2005).
\newblock {\em The grammar of society: The nature and dynamics of social norms}.
\newblock Cambridge University Press.

\bibitem[\protect\citeauthoryear{Cui and Wu}{Cui and Wu}{2013}]{cui2013impact}
Cui, P.-B. and Z.-X. Wu (2013).
\newblock Impact of conformity on the evolution of cooperation in the prisoner’s dilemma game.
\newblock {\em Physica A: Statistical Mechanics and its Applications\/}~{\em 392\/}(6), 1500--1509.

\bibitem[\protect\citeauthoryear{d'Adda, Dufwenberg, Passarelli, and Tabellini}{d'Adda et~al.}{2020}]{d2020social}
d'Adda, G., M.~Dufwenberg, F.~Passarelli, and G.~Tabellini (2020).
\newblock Social norms with private values: Theory and experiments.
\newblock {\em Games and Economic Behavior\/}~{\em 124}, 288--304.

\bibitem[\protect\citeauthoryear{Fang, Xu, Perc, and Chen}{Fang et~al.}{2019}]{fang2019effect}
Fang, Y., H.~Xu, M.~Perc, and S.~Chen (2019).
\newblock The effect of conformists’ behavior on cooperation in the spatial public goods game.
\newblock In {\em International Conference on Group Decision and Negotiation}, pp.\  137--145. Springer.

\bibitem[\protect\citeauthoryear{Hardin}{Hardin}{1968}]{hardin1968tragedy}
Hardin, G. (1968).
\newblock The tragedy of the commons: the population problem has no technical solution; it requires a fundamental extension in morality.
\newblock {\em science\/}~{\em 162\/}(3859), 1243--1248.

\bibitem[\protect\citeauthoryear{Harring and Krockow}{Harring and Krockow}{2021}]{harring2021social}
Harring, N. and E.~M. Krockow (2021).
\newblock The social dilemmas of climate change and antibiotic resistance: an analytic comparison and discussion of policy implications.
\newblock {\em Humanities and Social Sciences Communications\/}~{\em 8\/}(1).

\bibitem[\protect\citeauthoryear{Hu, Guo, Geng, and Shi}{Hu et~al.}{2019}]{hu2019effect}
Hu, K., H.~Guo, Y.~Geng, and L.~Shi (2019).
\newblock The effect of conformity on the evolution of cooperation in multigame.
\newblock {\em Physica A: Statistical Mechanics and its Applications\/}~{\em 516}, 267--272.

\bibitem[\protect\citeauthoryear{Huang, Li, and Jiang}{Huang et~al.}{2023}]{huang2023dual}
Huang, C., Y.~Li, and L.~Jiang (2023).
\newblock Dual effects of conformity on the evolution of cooperation in social dilemmas.
\newblock {\em Physical Review E\/}~{\em 108\/}(2), 024123.

\bibitem[\protect\citeauthoryear{Javarone, Antonioni, and Caravelli}{Javarone et~al.}{2016}]{javarone2016conformity}
Javarone, M.~A., A.~Antonioni, and F.~Caravelli (2016).
\newblock Conformity-driven agents support ordered phases in the spatial public goods game.
\newblock {\em Europhysics Letters\/}~{\em 114\/}(3), 38001.

\bibitem[\protect\citeauthoryear{Lin, Huang, Dai, and Yang}{Lin et~al.}{2020}]{lin2020evolutionary}
Lin, J., C.~Huang, Q.~Dai, and J.~Yang (2020).
\newblock Evolutionary game dynamics of combining the payoff-driven and conformity-driven update rules.
\newblock {\em Chaos, Solitons \& Fractals\/}~{\em 140}, 110146.

\bibitem[\protect\citeauthoryear{Nowak, Tarnita, and Antal}{Nowak et~al.}{2010}]{nowak2010evolutionary}
Nowak, M.~A., C.~E. Tarnita, and T.~Antal (2010).
\newblock Evolutionary dynamics in structured populations.
\newblock {\em Philosophical Transactions of the Royal Society B: Biological Sciences\/}~{\em 365\/}(1537), 19--30.

\bibitem[\protect\citeauthoryear{Pena, Volken, Pestelacci, and Tomassini}{Pena et~al.}{2009}]{pena2009conformity}
Pena, J., H.~Volken, E.~Pestelacci, and M.~Tomassini (2009).
\newblock Conformity hinders the evolution of cooperation on scale-free networks.
\newblock {\em Physical Review E—Statistical, Nonlinear, and Soft Matter Physics\/}~{\em 80\/}(1), 016110.

\bibitem[\protect\citeauthoryear{Perc, G{\'o}mez-Gardenes, Szolnoki, Flor{\'\i}a, and Moreno}{Perc et~al.}{2013}]{perc2013evolutionary}
Perc, M., J.~G{\'o}mez-Gardenes, A.~Szolnoki, L.~M. Flor{\'\i}a, and Y.~Moreno (2013).
\newblock Evolutionary dynamics of group interactions on structured populations: a review.
\newblock {\em Journal of the royal society interface\/}~{\em 10\/}(80), 20120997.

\bibitem[\protect\citeauthoryear{Santos, Santos, and Pacheco}{Santos et~al.}{2008}]{santos2008social}
Santos, F.~C., M.~D. Santos, and J.~M. Pacheco (2008).
\newblock Social diversity promotes the emergence of cooperation in public goods games.
\newblock {\em Nature\/}~{\em 454\/}(7201), 213--216.

\bibitem[\protect\citeauthoryear{Selten}{Selten}{1990}]{selten1990bounded}
Selten, R. (1990).
\newblock Bounded rationality.
\newblock {\em Journal of Institutional and Theoretical Economics (JITE)/Zeitschrift f{\"u}r die gesamte Staatswissenschaft\/}~{\em 146\/}(4), 649--658.

\bibitem[\protect\citeauthoryear{Simon}{Simon}{1955}]{simon1955behavioral}
Simon, H.~A. (1955).
\newblock A behavioral model of rational choice.
\newblock {\em The quarterly journal of economics\/}, 99--118.

\bibitem[\protect\citeauthoryear{Sunstein}{Sunstein}{2019}]{sunstein2019conformity}
Sunstein, C.~R. (2019).
\newblock Conformity: The power of social influences.
\newblock In {\em Conformity}. New York University Press.

\bibitem[\protect\citeauthoryear{Szolnoki and Perc}{Szolnoki and Perc}{2015}]{szolnoki2015conformity}
Szolnoki, A. and M.~Perc (2015).
\newblock Conformity enhances network reciprocity in evolutionary social dilemmas.
\newblock {\em Journal of The Royal Society Interface\/}~{\em 12\/}(103), 20141299.

\bibitem[\protect\citeauthoryear{Szolnoki and Perc}{Szolnoki and Perc}{2016}]{szolnoki2016leaders}
Szolnoki, A. and M.~Perc (2016).
\newblock Leaders should not be conformists in evolutionary social dilemmas.
\newblock {\em Scientific Reports\/}~{\em 6\/}(1), 23633.

\bibitem[\protect\citeauthoryear{Szolnoki, Wang, and Perc}{Szolnoki et~al.}{2012}]{szolnoki2012wisdom}
Szolnoki, A., Z.~Wang, and M.~Perc (2012).
\newblock Wisdom of groups promotes cooperation in evolutionary social dilemmas.
\newblock {\em Scientific Reports\/}~{\em 2\/}(1), 576.

\bibitem[\protect\citeauthoryear{Tomassini and Antonioni}{Tomassini and Antonioni}{2020}]{tomassini2020public}
Tomassini, M. and A.~Antonioni (2020).
\newblock Public goods games on coevolving social network models.
\newblock {\em Frontiers in Physics\/}~{\em 8}, 58.

\bibitem[\protect\citeauthoryear{Traulsen and Glynatsi}{Traulsen and Glynatsi}{2023}]{traulsen2023future}
Traulsen, A. and N.~E. Glynatsi (2023).
\newblock The future of theoretical evolutionary game theory.
\newblock {\em Philosophical Transactions of the Royal Society B\/}~{\em 378\/}(1876), 20210508.

\bibitem[\protect\citeauthoryear{Van~Lange and Rand}{Van~Lange and Rand}{2022}]{van2022human}
Van~Lange, P.~A. and D.~G. Rand (2022).
\newblock Human cooperation and the crises of climate change, covid-19, and misinformation.
\newblock {\em Annual Review of Psychology\/}~{\em 73\/}(1), 379--402.

\bibitem[\protect\citeauthoryear{Von~Neumann and Morgenstern}{Von~Neumann and Morgenstern}{1953}]{von1944theory}
Von~Neumann, J. and O.~Morgenstern (1953).
\newblock Theory of games and economic behavior, princeton.

\bibitem[\protect\citeauthoryear{Wang and Sun}{Wang and Sun}{2023}]{wang2023public}
Wang, C. and C.~Sun (2023).
\newblock Public goods game across multilayer populations with different densities.
\newblock {\em Chaos, Solitons \& Fractals\/}~{\em 168}, 113154.

\bibitem[\protect\citeauthoryear{Wang and Chen}{Wang and Chen}{2003}]{wang2003complex}
Wang, X.~F. and G.~Chen (2003).
\newblock Complex networks: small-world, scale-free and beyond.
\newblock {\em IEEE circuits and systems magazine\/}~{\em 3\/}(1), 6--20.

\bibitem[\protect\citeauthoryear{Watts and Strogatz}{Watts and Strogatz}{1998}]{watts1998collective}
Watts, D.~J. and S.~H. Strogatz (1998).
\newblock Collective dynamics of ‘small-world’networks.
\newblock {\em nature\/}~{\em 393\/}(6684), 440--442.

\bibitem[\protect\citeauthoryear{Yang and Tian}{Yang and Tian}{2017}]{yang2017enhancement}
Yang, H.-X. and L.~Tian (2017).
\newblock Enhancement of cooperation through conformity-driven reproductive ability.
\newblock {\em Chaos, Solitons \& Fractals\/}~{\em 103}, 159--162.

\bibitem[\protect\citeauthoryear{Zhang, Fan, and Luo}{Zhang et~al.}{2017}]{zhang2017emergence}
Zhang, Y., R.~Fan, and M.~Luo (2017).
\newblock Emergence of group cooperation in public goods game on regular small-world network.
\newblock {\em Wuhan University Journal of Natural Sciences\/}~{\em 22\/}(6), 529--534.

\end{thebibliography}
\bibliographystyle{chicago}

\section{Authorship contributions}
\textbf{Study conception and design}: RManfredi, RMastrandrea.\\
\textbf{Simulations}: RManfredi. \\
\textbf{Interpretation of results}: all authors. \\
\textbf{Draft manuscript preparation}: RManfredi. \\
\textbf{Manuscript revision}: all authors. \\

\appendix
\section{Appendix} \label{appendix}

\subsection{Payoff dynamics in defectors' clusters in the circular lattice} \label{Appendix_A}

\begin{figure}
    \centering
    \includegraphics[width=0.6\linewidth]{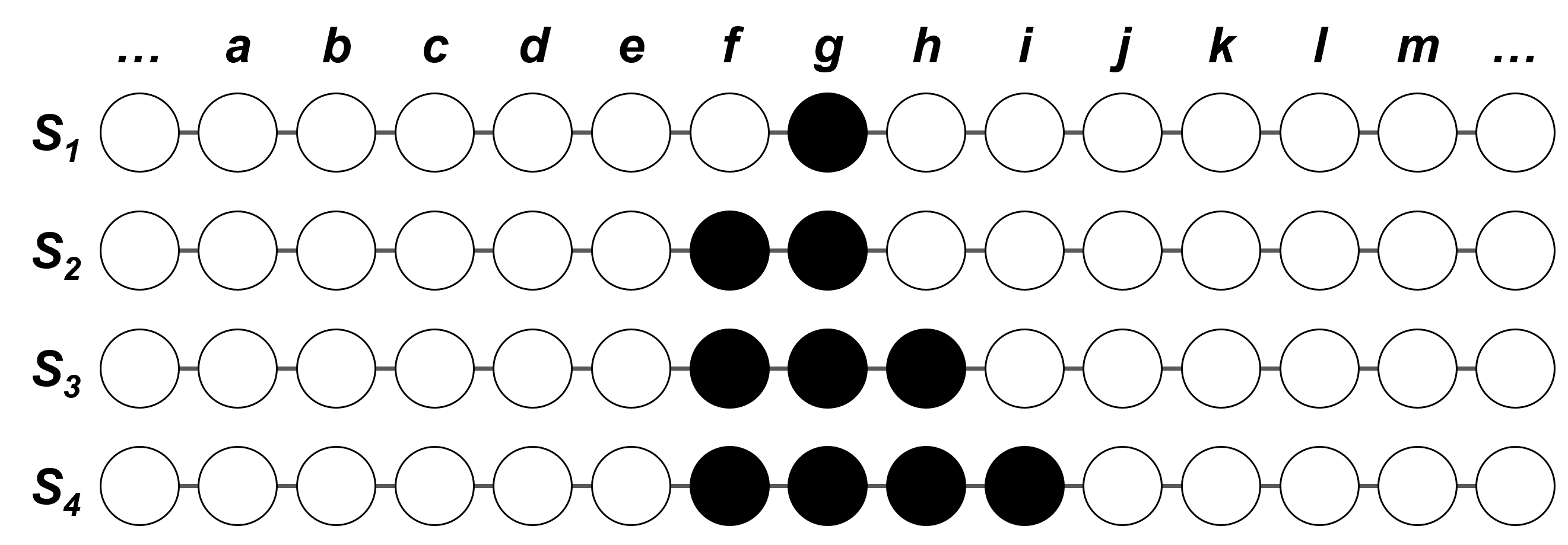}
    \caption{Four different scenario with increasing number of defectors (black circles) in a network of cooperators.}
    \label{fig:CL_payoff_C_D}
\end{figure}

In the circular lattice, the presence of a defector or a group of defectors can persist even when they are in a clear minority compared to the rest of the population, due to the difference in payoff between them and their cooperating neighbors, leading to the formation of clusters or individual “stubborn” nodes, capable of maintaining their strategy virtually unchanged (in the absence of conformity or random mutations).\\
\noindent Let's consider the four scenarios shown in Figure \ref{fig:CL_payoff_C_D}. For each case, we compute the payoff of defectors and only the payoff of cooperators with the highest value among the agents that each D could imitate. In the first scenario, $S_1$, a single defector (agent g, black circle) is inserted into a network of cooperators (white circles). We can compute the total payoff obtained from all public goods games played for the defector agent $g$ and the cooperators $e$ and $i$. In this case, the probability that a defector will imitate any cooperator within its own neighborhood is zero for any value of the enhancement factor $r \leq \frac{25}{2}$.

\begin{align*}
    &\pi^D_g = 4r \\
    &\pi^C_e = \pi^C_i = \frac{22}{5}r - 5 \\
    &P_{(g \rightarrow e,i)} = 0, \quad\forall r \leq \frac{25}{2}
\end{align*}

\noindent The gap between the payoffs of the two types of agent tends to gradually decrease as the number of defectors increases. This effect can already be observed with a single additional defector (case $S_2$).  Here, the defectors $g$ and $f$ will not imitate, respectively, cooperators $i$ and $d$ for any value of the enhancement factor $r \leq \frac{25}{4}$.

\begin{align*}
    &\pi^D_g = \pi^D_f = \frac{16}{5}r \\
    &\pi^C_d = \pi^C_i = 4r - 5 \\
    &P_{(g \rightarrow i)} = P_{(f \rightarrow d)} = 0, \quad \forall r\leq \frac{25}{4}
\end{align*}

\noindent If we increase the number of defectors within the same neighborhood, progressively smaller values of $r$ will be necessary to obtain a positive probability of imitation. For example, in the case of three adjacent defectors ($S_3$), defector $g$ will not imitate cooperators $e$ or $i$ until $r\leq \frac{25}{4}$, while the defectors $f$ and $h$ will never imitate cooperators $d$ or $j$ for $r\leq \frac{25}{6}$.

\begin{align*}
    & \pi^D_g = \frac{12}{5}r \\
    &\pi^D_f = \pi^D_h = \frac{13}{5}r \\
    &\pi^C_e = \pi^C_i = \frac{16}{5}r - 5 \\
    &\pi^C_d = \pi^C_j = \frac{19}{5}r - 5 \\
    &P_{(g \rightarrow e,i)} = 0, \quad \forall r\leq \frac{25}{4} \\
    &P_{(f,h \rightarrow d,j)} = 0, \quad \forall r\leq \frac{25}{6}
\end{align*}

\noindent Similarly, for the case $S_4$, we can see how the threshold value of $r$ decreases further. Now defectors $f$ and $i$ have a positive probability of imitating agents $d$ and $k$ if $r > \frac{25}{8}$.

\begin{align*}
    &\pi^D_g = \pi^D_h =\frac{9}{5}r \\
    &\pi^D_f = \pi^D_i = \frac{11}{5}r \\
    &\pi^C_e = \pi^C_j = 3r - 5 \\
    &\pi^C_d = \pi^C_k = \frac{19}{5}r - 5 \\
    &P_{(g,h \rightarrow e,j)} = 0, \quad \forall r\leq \frac{25}{6} \\
    &P_{(f,i \rightarrow d,k)} = 0, \quad \forall r\leq \frac{25}{8}
\end{align*}

\subsection{Numerical results} \label{Appendix_B}
Table \ref{all_results_numerical}  shows the numerical results obtained from the simulations described in Section \ref{results_ch1} and represented in Figures \ref{fig:CL_Overview}, \ref{fig:SW_overview} and \ref{fig:SF_overview}. For each of the topologies employed (circular lattice, small-world, scale-free), four scenarios were tested by varying the enhancement factor $r$. The first (“NoSN”) corresponds to the standard spatial PGG in the absence of conformity. The latter is then introduced by assigning a parameter $\alpha$ to each agent, based on three different Beta distributions. \\

\begin{table}
\centering
\begin{tabular}{ccccc} \toprule
    \textbf{r} & \textbf{{NoSN}} & \textbf{{$\beta$(1,1)}} & \textbf{{$\beta$(1,10)}} & \textbf{{$\beta$(10,1)}} \\ \midrule
    \multicolumn{5}{c}{Circular lattice} \\ \midrule
    \textbf{1} & 0.000 & 0.000 & 0.000 & 0.000 \\
    & (0.000) & (0.000) & (0.000) & (0.000) \\
    \textbf{2} & 0.000 & 0.000 & 0.000 & 0.000 \\
    & (0.000) & (0.000) & (0.000) & (0.000) \\
    \textbf{3} & 0.695 & 0.869 & 1.000 & 0.814 \\
    & (0.190) & (0.196) & (0.374) & (0.455) \\
    \textbf{4} & 0.981 & 1.000 & 1.000 & 1.000 \\
    & (0.000) & (0.000) & (0.007) & (0.007) \\
    \textbf{5} & 0.976 & 1.000 & 1.000 & 1.000 \\
    & (0.000) & (0.000) & (0.000) & (0.004) \\
    \midrule
    \multicolumn{5}{c}{Small-world} \\ \midrule
    \textbf{1} & 0.000 & 0.000 & 0.000 & 0.000 \\
    & (0.000) & (0.000) & (0.000) & (0.000) \\
    \textbf{2} & 0.000 & 0.000 & 0.000 & 0.000 \\
    & (0.000) & (0.000) & (0.000) & (0.000) \\
    \textbf{3} & 0.011 & 0.000 & 0.033 & 0.000 \\
    & (0.017) & (0.000) & (0.181) & (0.010) \\
    \textbf{4} & 0.916 & 1.000 & 0.989 & 1.000 \\
    & (0.042) & (0.000) & (0.105) & (0.009) \\
    \textbf{5} & 0.990 & 1.000 & 1.000 & 1.000 \\
    & (0.006) & (0.000) & (0.000) & (0.000) \\
    \midrule
    \multicolumn{5}{c}{Scale-free} \\ \midrule
    \textbf{1} & 0.000 & 0.000 & 0.020 & 0.000 \\
    & (0.000) & (0.000) & (0.141) & (0.000) \\
    \textbf{2} & 0.000 & 0.000 & 0.200 & 0.000 \\
    & (0.000) & (0.000) & (0.401) & (0.000) \\
    \textbf{3} & 0.703 & 0.333 & 0.427 & 0.678 \\
    & (0.447) & (0.473) & (0.501) & (0.470) \\
    \textbf{4} & 1.000 & 0.714 & 0.487 & 1.000 \\
    & (0.012) & (0.453) & (0.496) & (0.000) \\
    \textbf{5} & 1.000 & 0.983 & 0.540 & 1.000 \\
    & (0.011) & (0.120) & (0.500) & (0.000) \\
    \midrule
\end{tabular}
\caption{Average final fraction of cooperators for each scenario tested in the three different topologies and for each value of $r$. The values in parentheses indicate the standard deviation and help identify scenarios in which conformity leads to bistability in the system.}
\label{all_results_numerical}
\end{table}

\subsection{Convergence patterns}
\label{Appendix_C}

\begin{figure}[bth]
    \centering
    \begin{subfigure}[t]{0.48\linewidth}
        \centering
        \includegraphics[width=\linewidth]{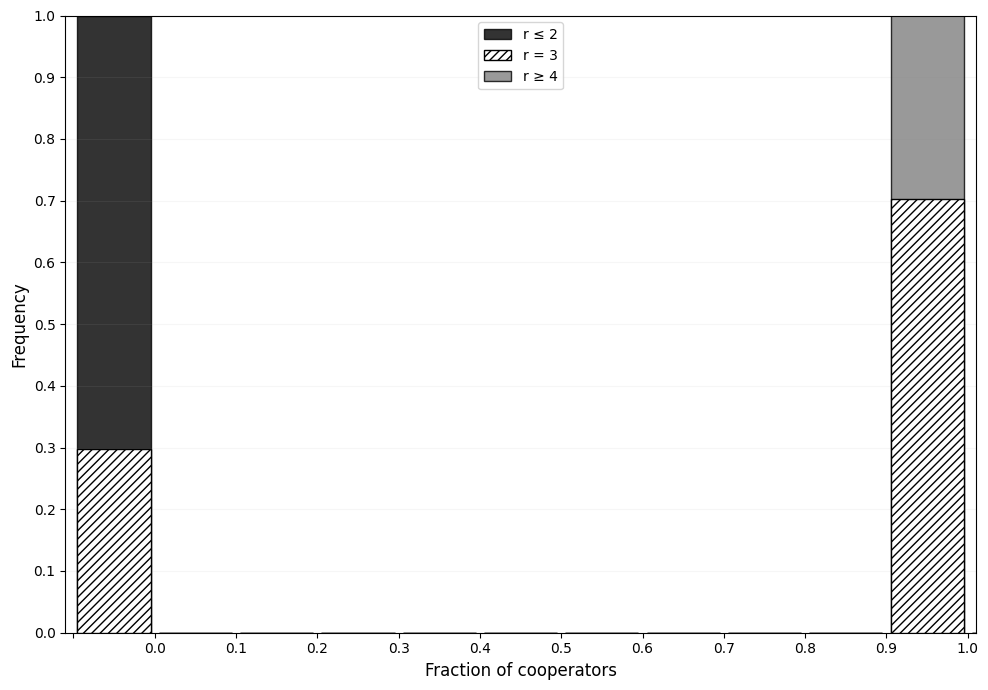}
        \caption{NoSN}
        \label{Conv_CL_NoSN_All}
    \end{subfigure}
    \hfill
    \begin{subfigure}[t]{0.48\linewidth}
        \centering
        \includegraphics[width=\linewidth]{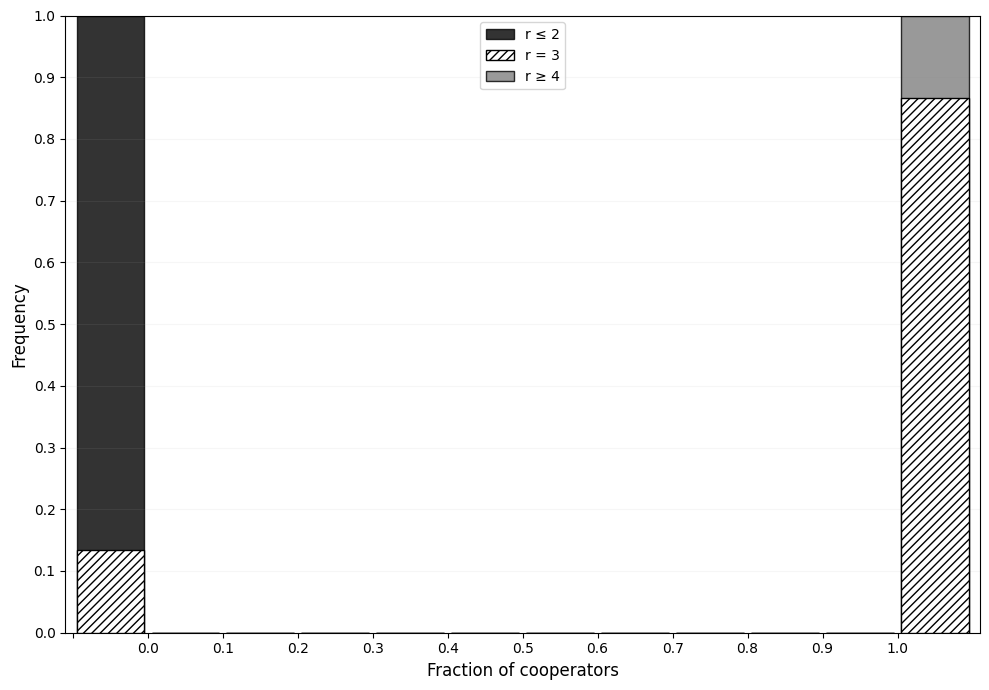}
        \caption{$\beta$(1,1)}
        \label{Conv_CL_B11_All}
    \end{subfigure}
    \hfill
    \begin{subfigure}[t]{0.48\linewidth}
        \centering
        \includegraphics[width=\linewidth]{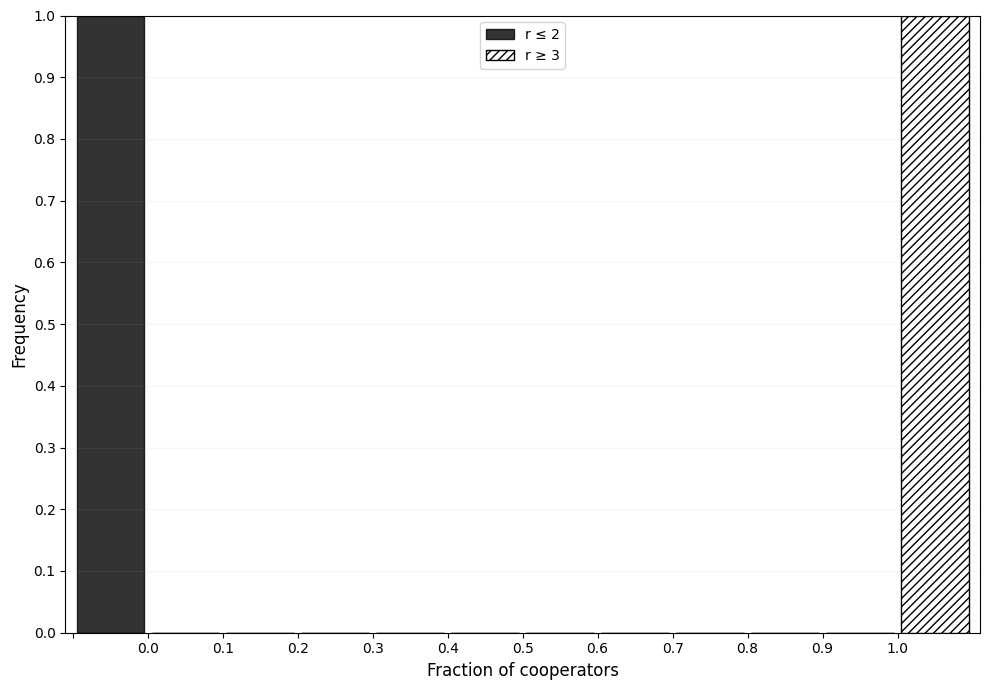}
        \caption{$\beta$(1,10)}
        \label{Conv_CL_B110_All}
    \end{subfigure}
    \hfill
    \begin{subfigure}[t]{0.48\linewidth}
        \centering
        \includegraphics[width=\linewidth]{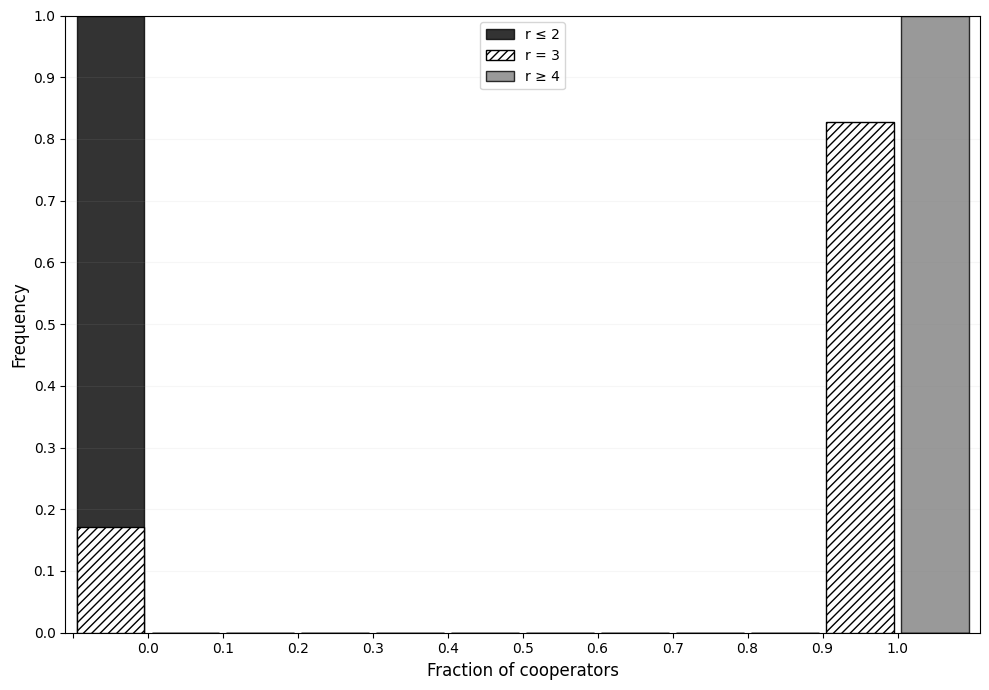}
        \caption{$\beta$(10,1)}
        \label{Conv_CL_B101_All}
    \end{subfigure}
    \caption{Convergence patterns in the circular lattice for the four scenarios tested.}
    \label{Conv_CL_All}
\end{figure}

\begin{figure}[bth]
    \centering
    \begin{subfigure}[t]{0.48\linewidth}
        \centering
        \includegraphics[width=\linewidth]{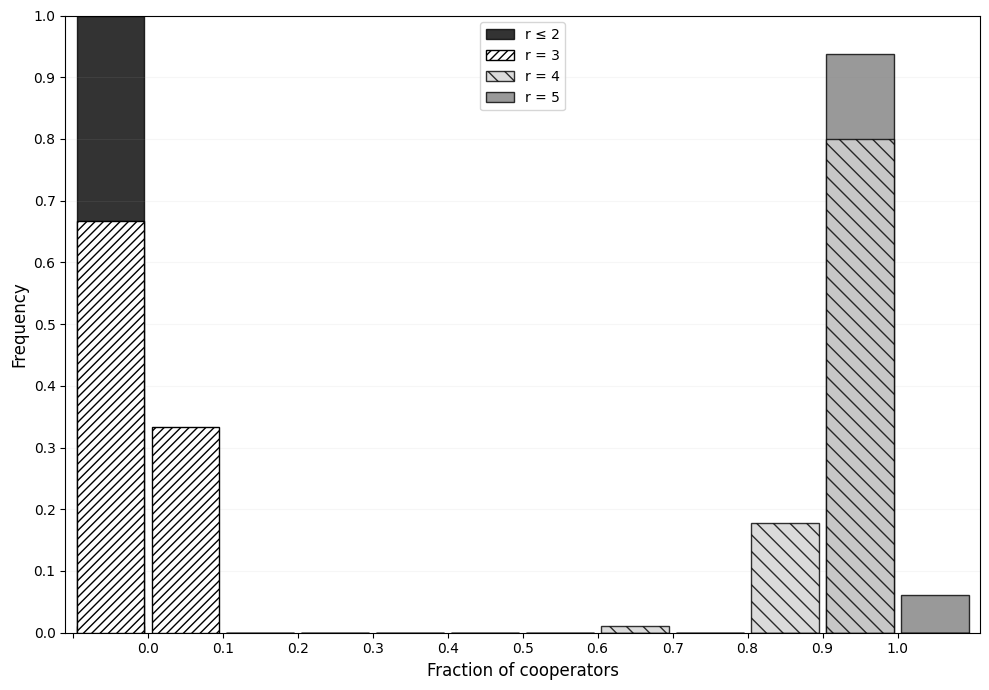}
        \caption{NoSN}
        \label{Conv_SW_NoSN_All}
    \end{subfigure}
    \hfill
    \begin{subfigure}[t]{0.48\linewidth}
        \centering
        \includegraphics[width=\linewidth]{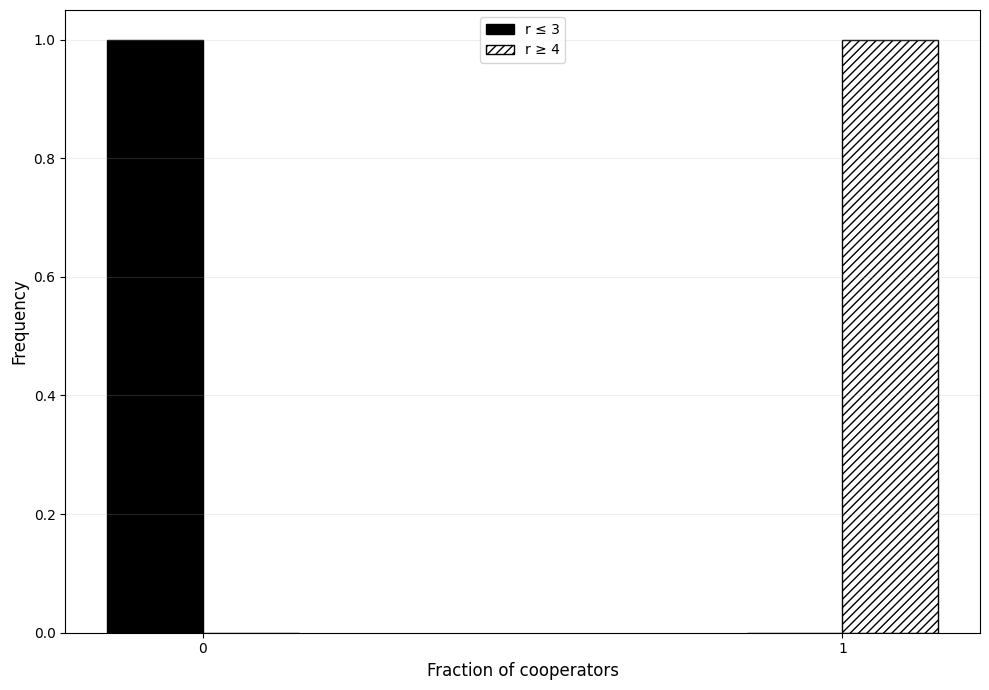}
        \caption{$\beta$(1,1)}
        \label{Conv_SW_B11_All}
    \end{subfigure}
    \hfill
    \begin{subfigure}[t]{0.48\linewidth}
        \centering
        \includegraphics[width=\linewidth]{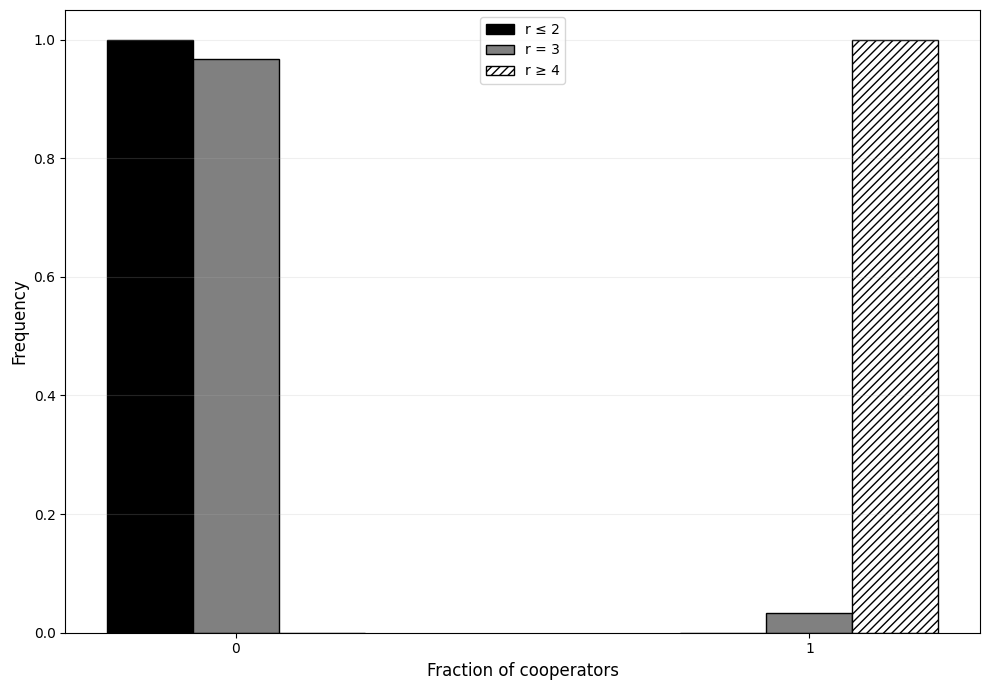}
        \caption{$\beta$(1,10)}
        \label{Conv_SW_B110_All}
    \end{subfigure}
    \hfill
    \begin{subfigure}[t]{0.48\linewidth}
        \centering
        \includegraphics[width=\linewidth]{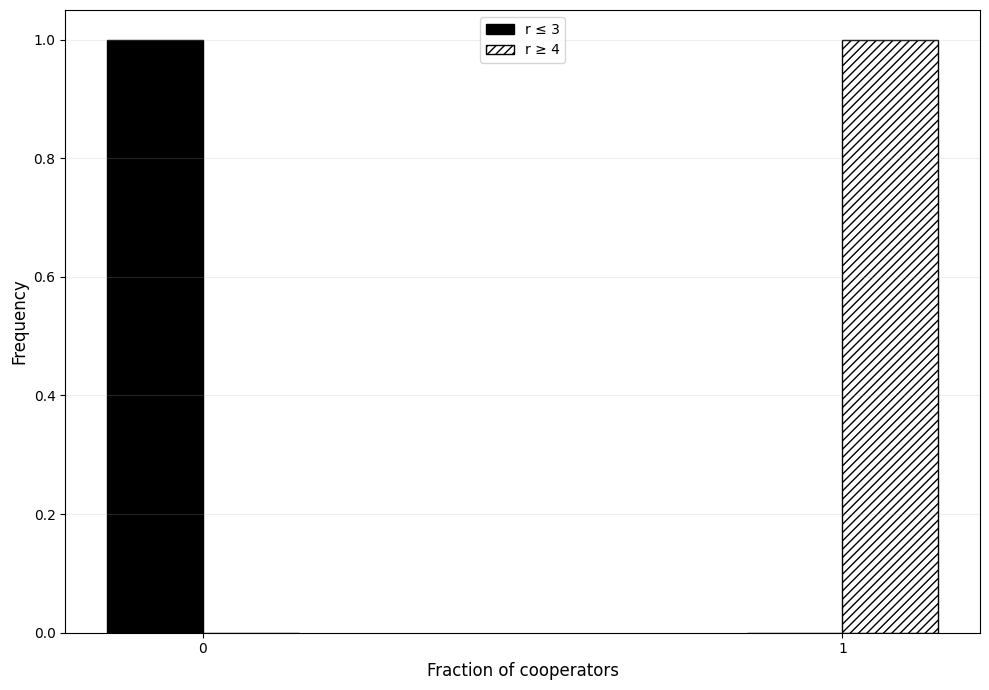}
        \caption{$\beta$(10,1)}
        \label{Conv_SW_B101_All}
    \end{subfigure}
    \caption{Convergence patterns in the small-world for the four scenarios tested.}
    \label{Conv_SW_All}
\end{figure}

\begin{figure}[bth]
    \centering
    \begin{subfigure}[t]{0.48\linewidth}
        \centering
        \includegraphics[width=\linewidth]{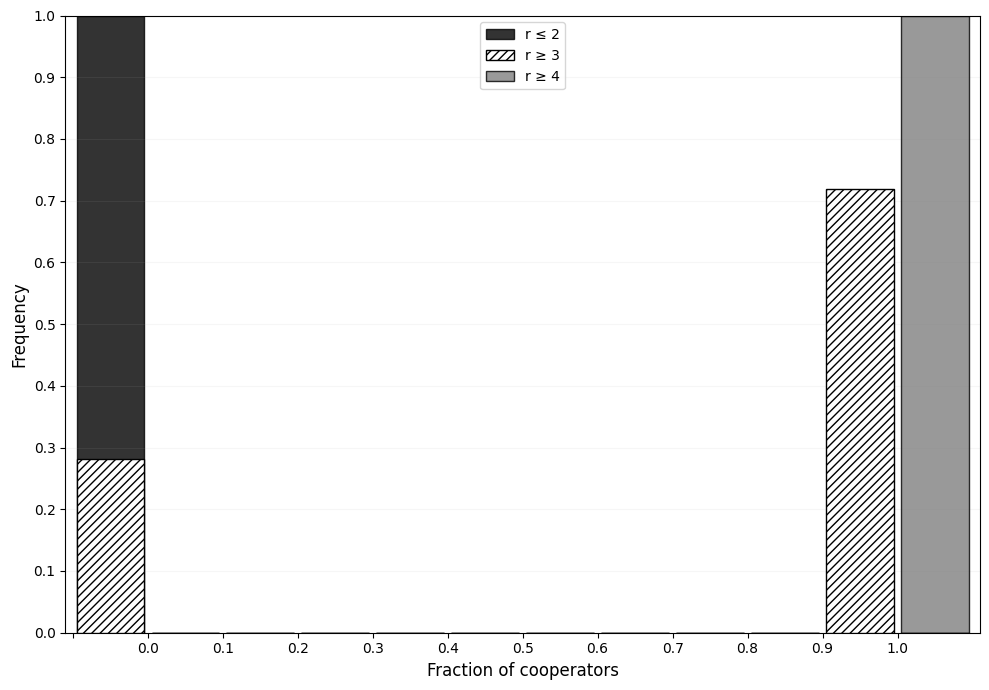}
        \caption{NoSN}
        \label{Conv_SF_NoSN_All}
    \end{subfigure}
    \hfill
    \begin{subfigure}[t]{0.48\linewidth}
        \centering
        \includegraphics[width=\linewidth]{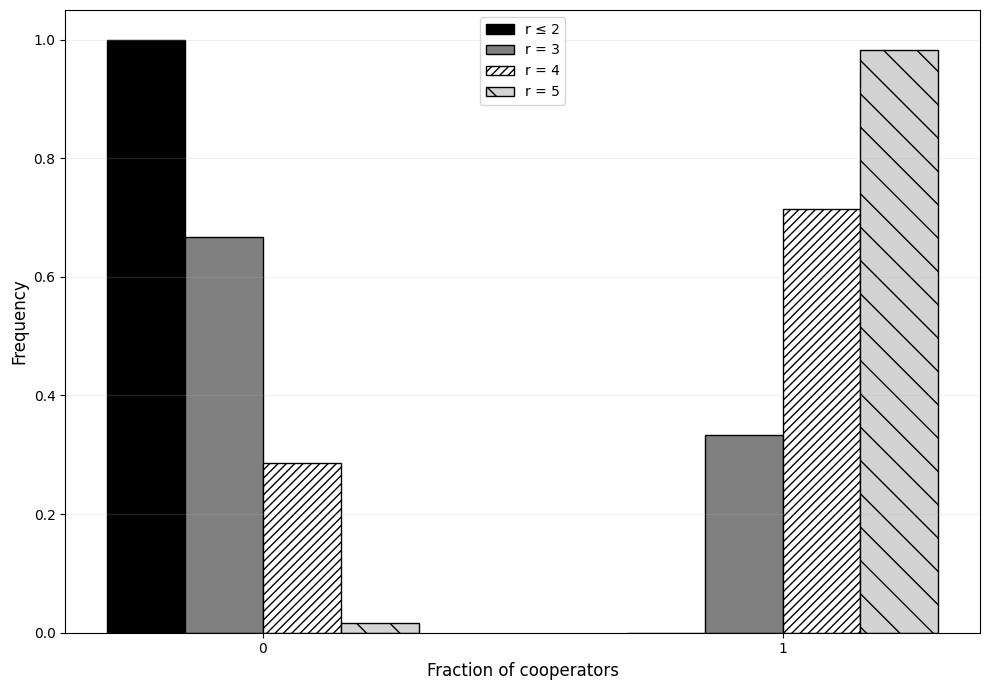}
        \caption{$\beta$(1,1)}
        \label{Conv_SF_B11_All}
    \end{subfigure}
    \hfill
    \begin{subfigure}[t]{0.48\linewidth}
        \centering
        \includegraphics[width=\linewidth]{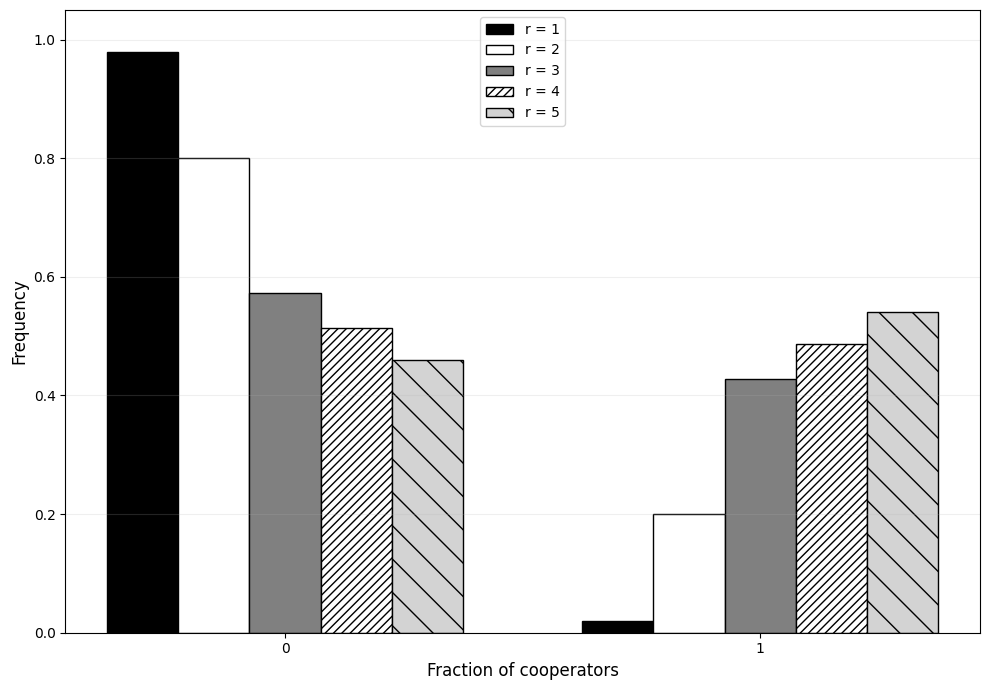}
        \caption{$\beta$(1,10)}
        \label{Conv_SF_B110_All}
    \end{subfigure}
    \hfill
    \begin{subfigure}[t]{0.48\linewidth}
        \centering
        \includegraphics[width=\linewidth]{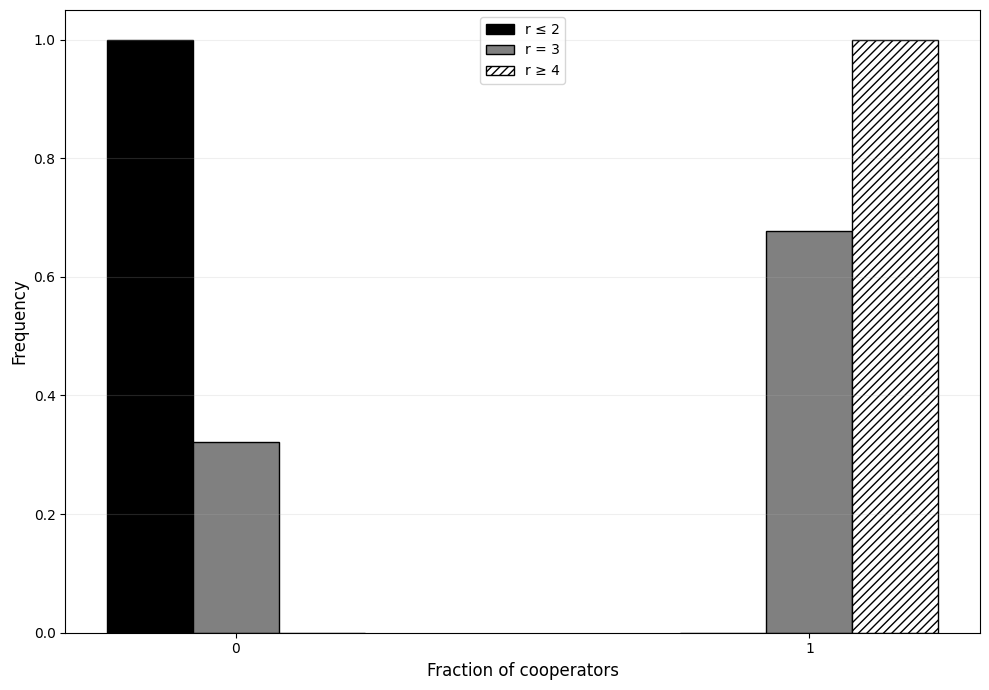}
        \caption{$\beta$(10,1)}
        \label{Conv_SF_B101_All}
    \end{subfigure}
    \caption{Convergence patterns in the scale-free for the four scenarios tested.}
    \label{Conv_SF_All}
\end{figure}

Figures \ref{Conv_CL_All}, \ref{Conv_SW_All}, and \ref{Conv_SF_All} show the convergence patterns in the three topologies employed and in each of the four scenarios tested. \\
\noindent In the circular lattice (Figure \ref{Conv_CL_All}), the system always converges to total defection for $r \leq 2$, regardless of the presence of conformity. When the latter is absent (Figure \ref{Conv_CL_NoSN_All}), cooperators never manage to completely invade the system. For $r = 3$, we observe an alternation between total defection and very high values of cooperation. For $r \geq 4$, the system is always dominated by cooperators, with a small immovable minority of defectors. This hard core of defectors is undermined when conformity is introduced. Even a weak presence is sufficient to achieve total cooperation for $r\geq4$, as we see in Figure \ref{Conv_CL_B101_All}. By further increasing the role of conformity in the agents' decision-making process, i.e., cases $\beta$(1,1) and $\beta$(1,10), the system becomes bistable. In the first case (Figure \ref{Conv_CL_B11_All}), for $r =3$ we observe convergence to both total defection and total cooperation, while for $r \geq 4$, cooperation manages to invade the system. In the case $\beta(1,10)$ (Figure \ref{Conv_CL_B110_All}), for $r \geq 3$ the cooperators always dominate.\\
\noindent Similar to the circular lattice, cooperation is always absent in the small-world for $r \leq 2$ (Figure \ref{Conv_SW_All}). In the absence of conformity (Figure \ref{Conv_SW_NoSN_All}), for $r = 3$ defectors tend to invade the system in most cases. However, less frequently, some cooperators manage to persist in a small minority. For $r > 4$, the situation is reversed and cooperation becomes the majority, although never total. By further increasing $r$, cooperation levels tend to remain high and, occasionally, the system may converge. With the introduction of conformity, convergence patterns vary significantly. For both $\beta(10,1)$ and $\beta(1,1)$ scenarios, its presence favors the achievement of total defection for $r \leq 3$ and total cooperation for $r\geq 4$ (Figures \ref{Conv_SW_B101_All} and \ref{Conv_SW_B11_All}). When the role of conformity becomes predominant (Figure \ref{Conv_SW_B110_All}), for $3 \leq r \leq 4$ the system becomes bistable. The latter disappears in favor of total cooperation only when $r = 5$.\\
\noindent Finally, Figure \ref{Conv_SF_All} shows the convergence patterns for the scale-free network. In the standard spatial PGG (Figure \ref{Conv_SF_NoSN_All}), for $r \leq2$ the system converges to total defection, while it always reaches full cooperation for $r \geq 4$. In the intermediate case $r = 3$, defectors can occasionally invade the system, but in most cases cooperation occupies almost the entire network. With the introduction of conformity (Figures \ref{Conv_SF_B11_All}, \ref{Conv_SF_B110_All}, \ref{Conv_SF_B101_All}), the convergence patterns change radically. In fact, the system has only two final equilibria: total cooperation or total defection. The occurrence of one or the other is determined by the value of the enhancement factor $r$ and the weight of conformity in the agents' decision-making process.\\

\end{document}